\RequirePackage{lineno}                                             
\documentclass[nim,nofootinbib,floatfix,twocolumn,showpacs,superscriptaddress]{revtex4}
\usepackage{graphicx,amsmath}
\usepackage{dcolumn}
\usepackage{bm}
\usepackage{rotating}
\usepackage{amsmath,amssymb,amsfonts}
\usepackage{color}	

\begin{document}

\title{The GlueX Central Drift Chamber: Design and Performance\\}


\def\groupcmu{\affiliation{Department of Physics, Carnegie Mellon University, Pittsburgh, Pennsylvania 15213, USA}}
\def\groupindi{\affiliation{Department of Physics, Indiana University, Bloomington, Indiana 47405 USA}}
\def\groupjlab{\affiliation{Thomas Jefferson National Accelerator Facility, Newport News, Virginia 23606, USA}}


\groupcmu


\author{Y.~Van~Haarlem}  \groupcmu
\author{C.~A.~Meyer}  \groupcmu
\author{F.~Barbosa} \groupjlab
\author{B.~Dey} \groupcmu
\author{D.~Lawrence} \groupjlab
\author{V.~Razmyslovich} \groupjlab
\author{E.~Smith} \groupjlab
\author{G.~Visser} \groupindi
\author{T.~Whitlatch} \groupjlab
\author{G.~Wilkin} \groupcmu
\author{B.~Zihlmann} \groupjlab




\begin{abstract}
Tests and studies concerning the design and performance of the GlueX Central Drift
 Chamber (CDC) are presented. A full-scale prototype was built to test and steer 
the mechanical and electronic design. Small scale prototypes were constructed to test 
for sagging and to do timing and resolution studies of the detector. These 
studies were used to choose the gas mixture and to program a Monte Carlo 
simulation that can predict the detector response in an external magnetic field. 
Particle 
identification and charge division possibilities were also investigated.
\end{abstract}

\maketitle

\section{Introduction}
The GlueX Central Drift Chamber (CDC) is part of the GlueX experiment in Hall D at 
Jefferson Lab (Newport News, VA). This
experiment aims to elucidate confinement in Quantum Chromo-dynamics, by searching for hybrid mesons that possess gluonic
degrees of freedom and exotic quantum numbers~\cite{Curtis}, and arise from
photo-production at 9~GeV~\cite{TD,DMS}. To achieve this goal, amplitude 
analysis  on numerous exclusive reactions must be carried out to
determine the quantum numbers of produced exotic mesons,
which decay into photons and charged particles. Clearly, an
overall hermetic detector with adequate resolution is essential,
and the CDC is a crucial subsystem.

Straw tubes are a well-established technology~\cite{dcd, Mark2, Mac, CDF, x1, TPC, Mark3, KEDR, ssc, amy, e706, x2, JETSET, x3, DELPHI, KL, streamer, AMPIR, x5, SCD1, SCD2, TAGX, e864, pickuppad, pamela, compass, atlas} that has long been recognized as an alternative to multi-wire tracking chambers. Due to their cylindrical geometry, straw tubes can have very low mass, which reduces multiple scattering of charged particles. Individual wire breakage does not affect neighboring wires because each straw is self-contained.  Another advantage of straw tubes is that they 
can support their own wire-tension and in this way the end-plates (plates that hold the straws) can be made much 
thinner compared to a conventional wire chamber, which also reduces multiple scattering.

The CDC will consist of 3500, 1.5~m long, straw tubes. The straws 
are mounted in 28 layers: 12 axial and 16 stereo; the placement of the straws is 
chosen to minimize tracking ambiguities. The inner and outer diameters of the 
chamber are 20
and 120~cm respectively. Gas plenums are constructed outside each end-plate. These plenums allow gas to be distributed to all the straws at the same time instead of feeding gas to straws individually. High voltage distribution boards and pre-amplifiers are mounted on the upstream gas-plenum, outside the gas volume. The output signals of the pre-amplifiers travel over a 20~m long cable to readout-electronics in VME crates.

\begin{figure}
\begin{tabular}{cc}
  \includegraphics[width=1.\linewidth]{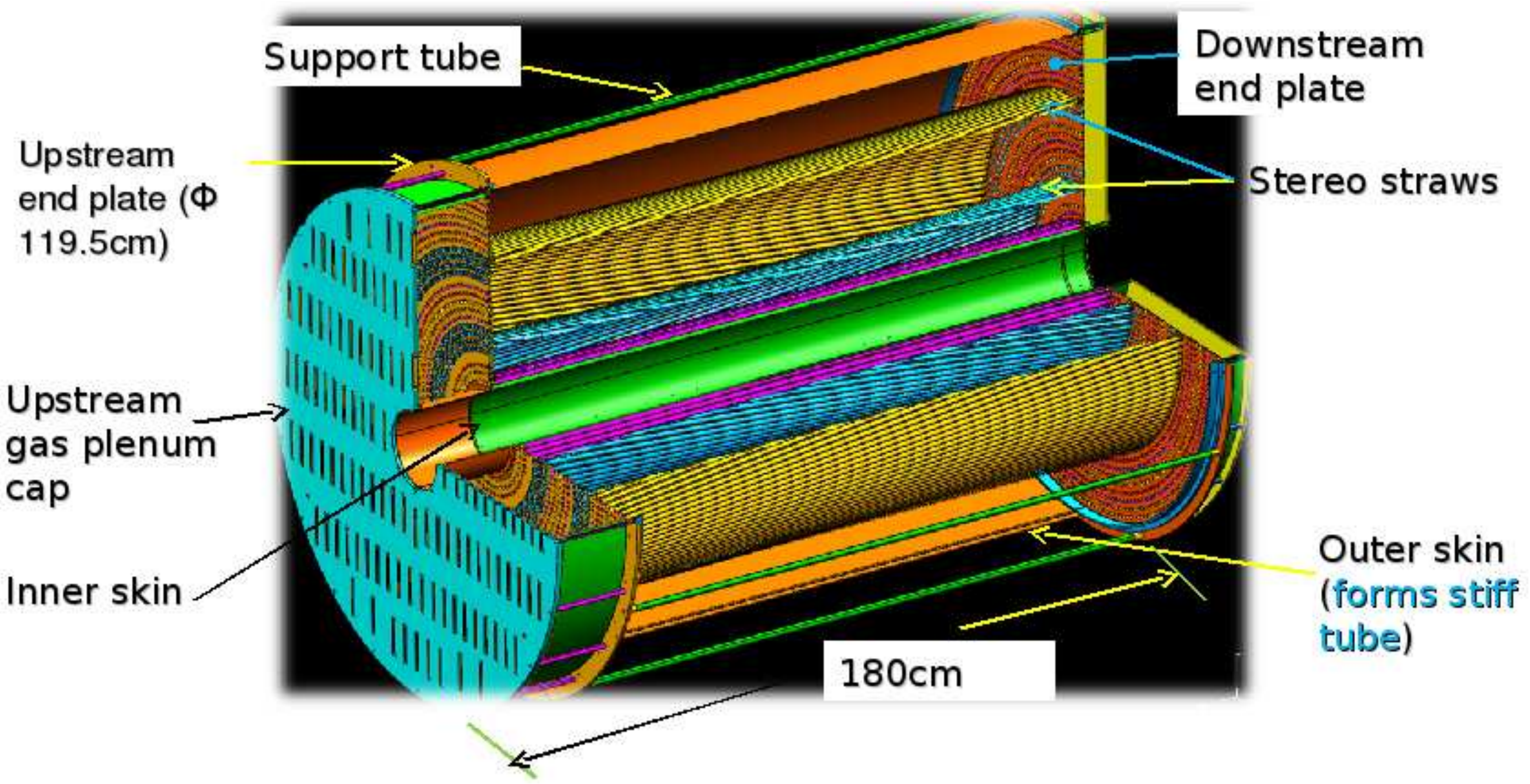}
\end{tabular}
\caption[]{\label{CDCfinal} Drawing of the final CDC design.}
\end{figure}

The CDC will be operated inside a 
 solenoid that generates a magnetic field of 2.24~T. 
The goal for position resolution is 150~$\mu$m transverse to the wire and 1.5~mm along the axis of the chamber. Full tracking will be achieved using both the CDC and the Forward Drift Chambers, which are located directly downstream from the CDC. The target will be located 25~cm upstream from the center of the CDC. 
The polar angular coverage is between 6 and 165$^{\circ}$. This detector is designed to enable measurements in energy loss ${\rm d}E/{\rm d}x$ needed for particle identification of pions, kaons, and protons with momenta 
lower than 450~MeV~\footnote{The speed of light ($c$) is set to one in this note.}. A drawing of the final CDC design is shown in Fig.~\ref{CDCfinal}.

First the straw tube design will be presented, with tests concerning sagging and radiation damage. Then 
the electronics and gas system is introduced, followed by studies to characterize 
the detector. These studies are used to construct a Monte Carlo simulation that 
describes the detector performance, followed by timing and resolution studies. The possibility of obtaining particle identification (PID) by measuring the energy loss of the particles in the straw volume is investigated using the Monte Carlo simulation. Possible improvement of position resolution along the wire by application of charge division is investigated, followed by final remarks and conclusions.

\section{Straw tube design}
\begin{figure}
\begin{tabular}{cc}
  \includegraphics[width=1.\linewidth]{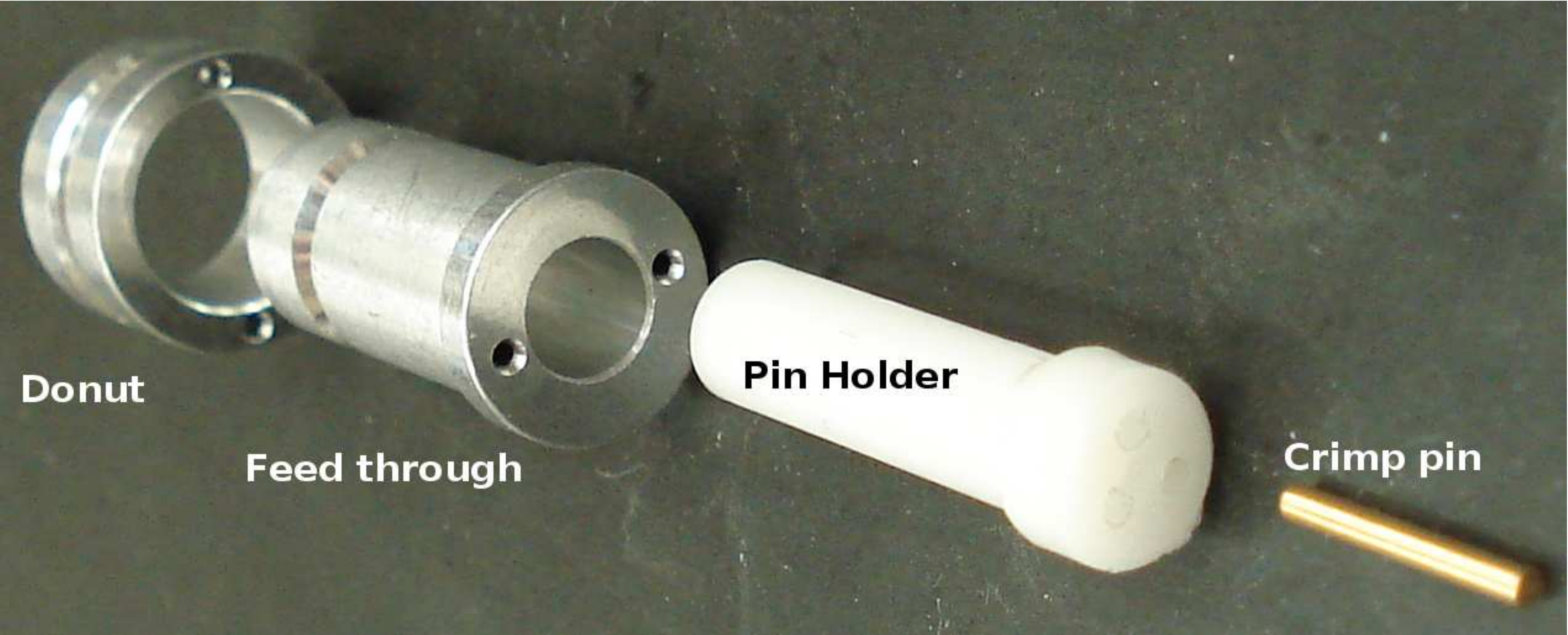}
\end{tabular}
\caption[]{\label{strawend} Picture showing the straw end assembly (upstream end). See text for more details.}
\end{figure}

The straws that will be used are 150~cm long, with an inner diameter of 1.6~cm. Fig.~\ref{strawend} shows a picture of the straw tube end assembly at the upstream end. The wires are 
held in place using a metallic crimp pin. The pin is inserted in a Noryl holder that 
has three extra holes to 
allow for gas flow into the straw. The pin holder is inserted into a feed-through that
is glued into the end-plate and into a donut. The donut is glued to the inside of the straw. At the upstream end of the chamber, the donut and feed-through are made of aluminum to provide an electrical ground connection between the end plate and the aluminum on the inside of the straw tube. At the downstream end, the donuts are made of Noryl to minimize the 
material along the particle tracks. Detailed specifications of the donut and feed-through can be found in~\cite{specsdonfeed}\footnote{GlueX documents mentioned in this note are public and can be found on http://argus.phys.uregina.ca/cgi-bin/public/DocDB/DocumentDatabase}.   
In following sections details about the wire and straws will be provided and the results of a sag and irradiation test are discussed. It is important that the straws do not sag because that causes the uniformity of the electric field to vanish. The irradiation test was performed to test the robustness of the aluminum layer inside the straw. If that layer gets damaged, it can cause noise or even a dead straw.

\subsection{Sense wire}
The sense wire material used in all tests and in the final design is gold plated tungsten wire manufactured by California Fine Wire\footnote{California Fine Wire, P.O. Box 446, Grover Beach, CA 93483-0446, http://www.calfinewire.com}. The wire has a total diameter of 20~$\mu$m. The tensile strength of this wire is 21,000~kg/cm$^2$. The wires in all test chambers are tensioned to 37~g, which is about half the yield point of 75~g. 
\begin{figure}
\begin{tabular}{cc}
  \includegraphics[width=0.65\linewidth]{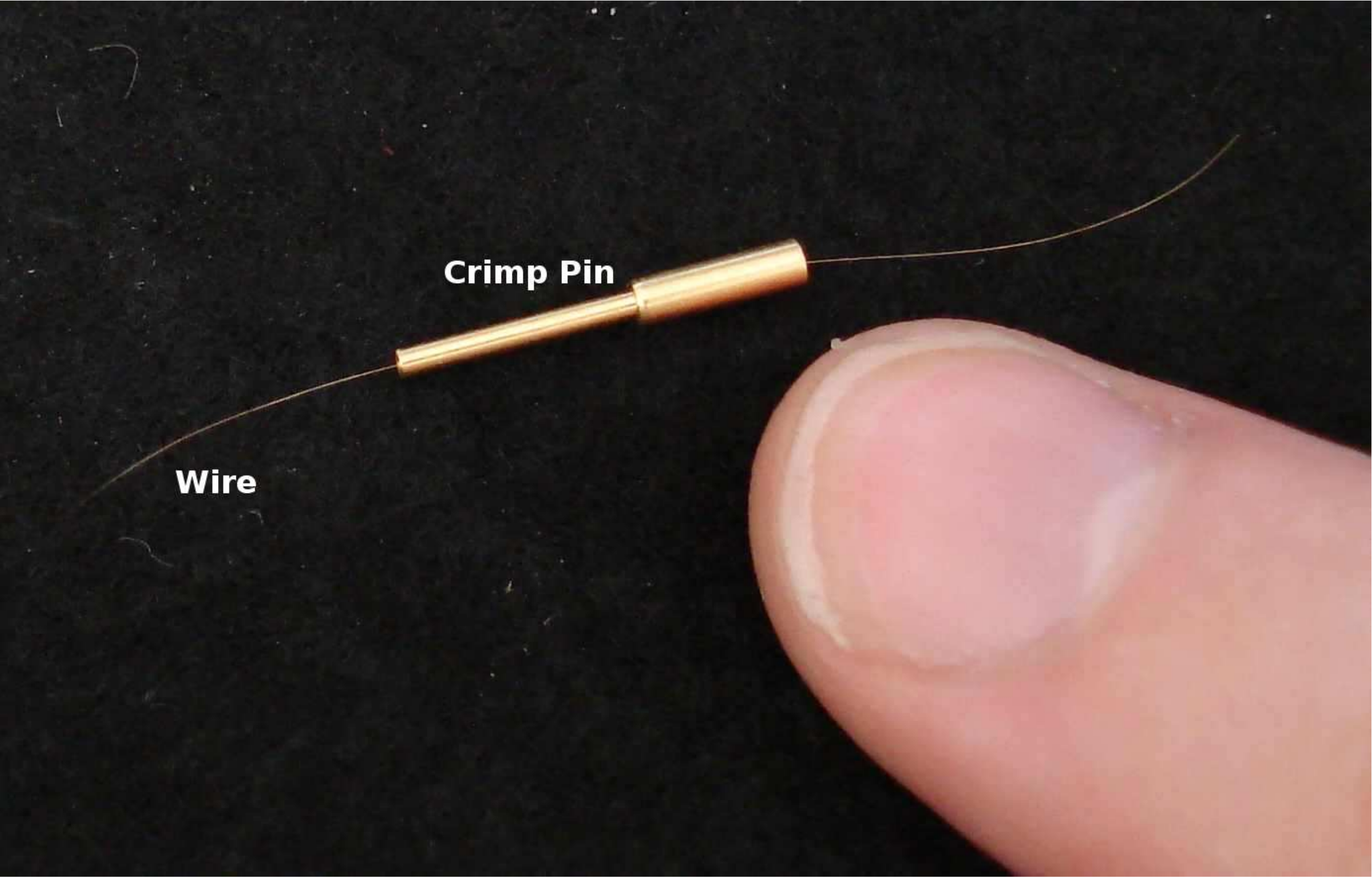}
\end{tabular}
\caption[]{\label{wirepin} Picture showing the wire used in the straw tubes together with a crimp pin (not the final design) that holds the wire in place.}
\end{figure}

\subsection{Straw}
A straw consists of several layers of thin (20-30~$\mu$m) mylar or kapton with a thin aluminum layer on the innermost one, which acts as a cathode. The straw tubes used for the CDC are 1.5~m long and have an inner radius of about 7.8~mm and a wall thickness of about 100~$\mu$m.       

\begin{figure}
\begin{tabular}{cc}
  \includegraphics[width=1.\linewidth]{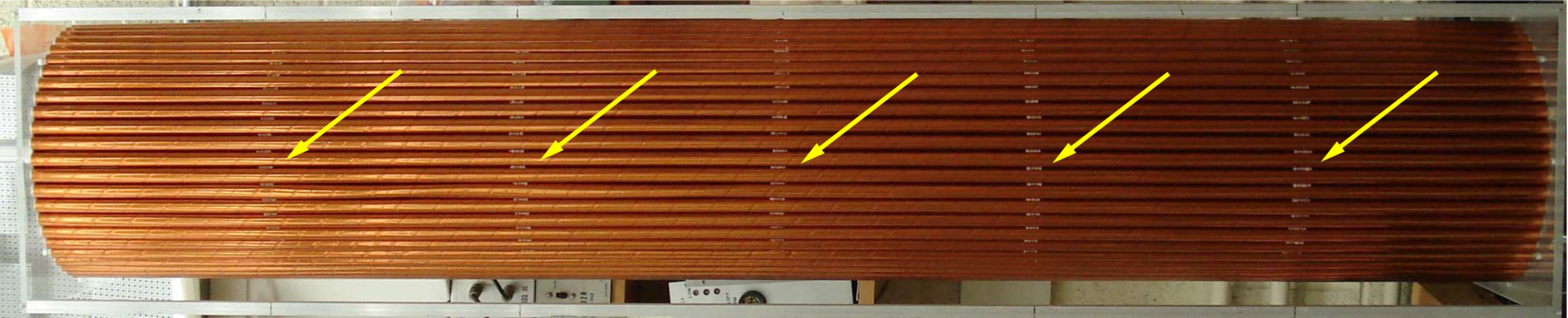}
\end{tabular}
\caption[]{\label{sagtest} Picture of a prototype used for sagging test. A full circle of 1.5~m long straws was put in place and spot-glued together. Rings of glue spots are visible and indicated by yellow arrows.}
\end{figure}

Individual straws of this length sag by several mm under their own weight. This sag can significantly distort the drift properties of the electrons. In order to avoid this, straws are mounted in layers of circles and spot glued together to prevent the sagging. This is tested by constructing a ring (diameter 25~cm) of 1.5~m long straws that are spot-glued together. The setup is shown in Fig.~\ref{sagtest}. No measurable sagging is observed.

\subsection{Radiation exposure test}
\subsubsection{Prototype}
To verify the stability and integrity of the aluminum layer at the inside surface of the straw an irradiation test was performed on a small scale prototype chamber with four different, 30~cm long, straw types. This prototype was fully operational and could be read out if desired.
The 4 straw types that were put into the testing prototype
 were a kapton straw with a $<$1~$\mu$m vapor-deposited aluminum layer on the 
inside produced by Stone Industrial\footnote{Stone Industrial, 9207 51st Ave., College Park, MD 20740-1910, http://www.stoneindustrial.com} (referred to as Stone-kapton), a mylar straw with a $<$1~$\mu$m vapor-deposited aluminum layer on the inside produced by Euclid Spiral Paper Tube Corporation\footnote{Euclid Spiral Paper Tube Corporation, 339 Mill Street, P.O. Box 458, Apple Creek, Ohio 44606, http://www.euclidspiral.com} (referred to as Euclid-mylar), and two straws consisting of 4 layers of 25~$\mu$m thick mylar
produced by Lamina Dielectrics Ltd\footnote{Lamina Dielectrics Ltd, Daux Road Billingshurst, West Sussex RH14 9SJ, United Kingdom, http://www.laminadielectrics.com}: one with 0.03~$\mu$m vapor-deposited aluminum on the inside mylar layer (referred to as Lamina-thin)
and one with a 12~$\mu$m thick aluminum foil glued on the inside mylar layer (referred to as Lamina-thick).
There were only two straw candidates considered for the final design: the Lamina-thick and Lamina-thin straw because the Euclid-mylar straw is not robust enough and the Stone-kapton straw is not as cost-effective as mylar.
\begin{figure}
\begin{tabular}{cc}
  \includegraphics[width=.755\linewidth]{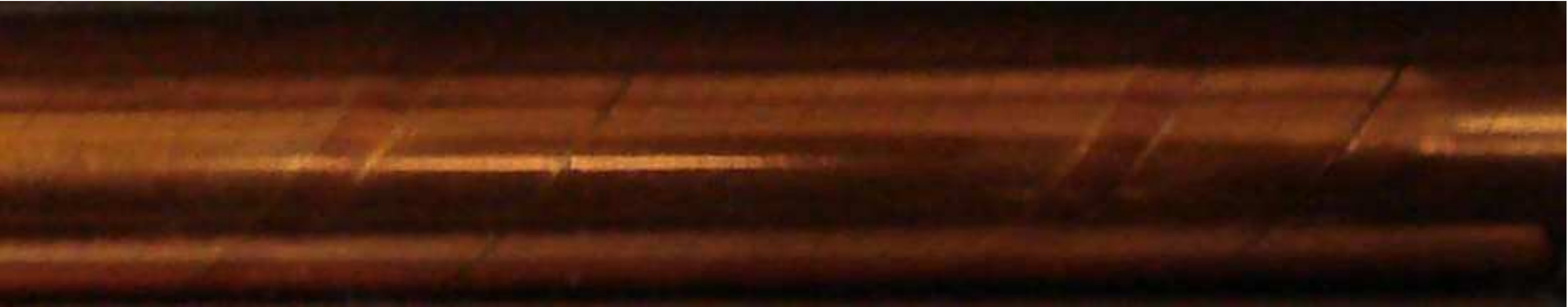}\\
  \includegraphics[width=.75\linewidth]{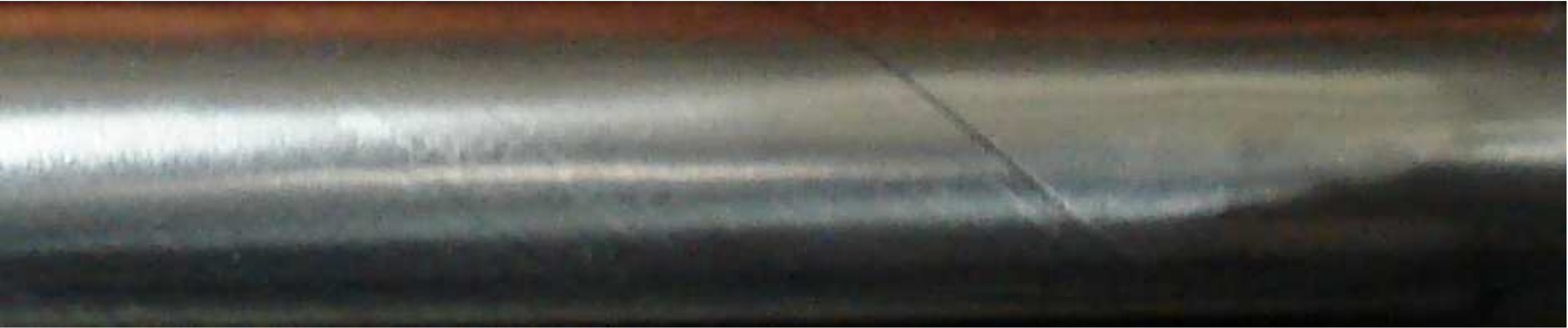}\\
  \includegraphics[width=.75\linewidth]{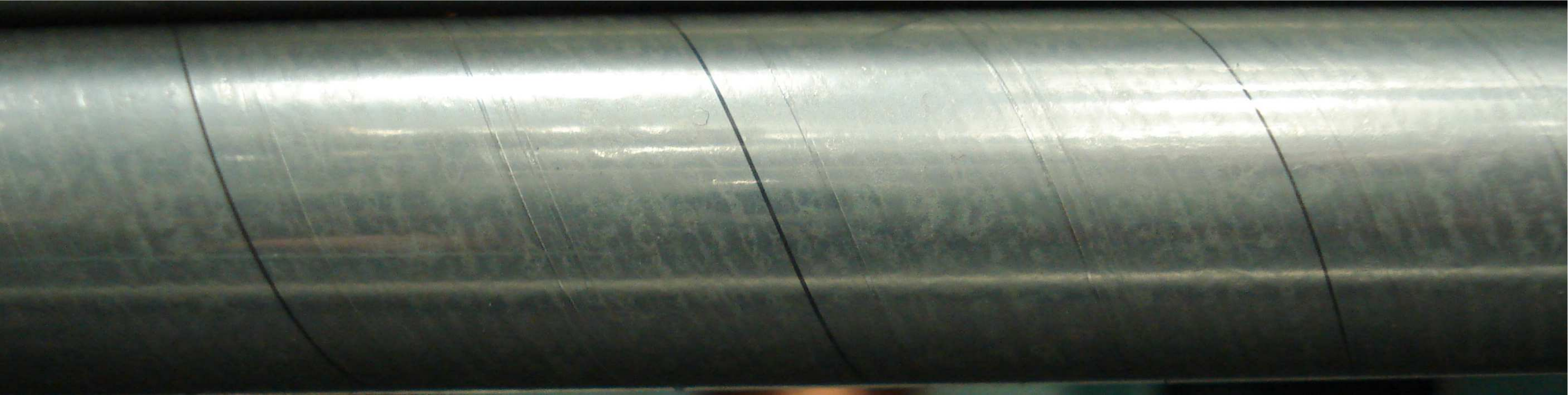}\\
  \includegraphics[width=.75\linewidth]{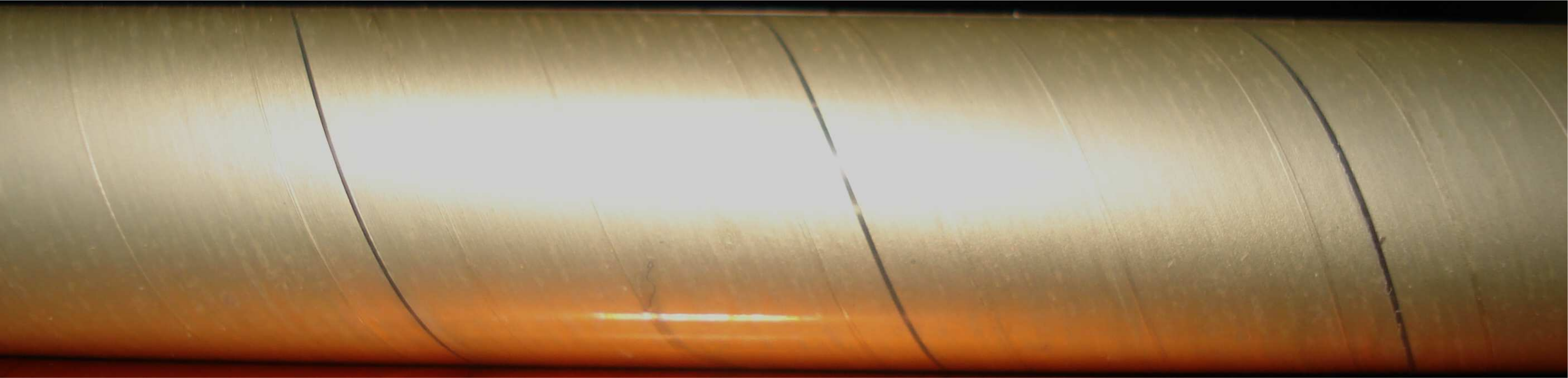}\\
\end{tabular}
\caption[]{\label{straws} Picture of the different straw types that were irradiated (not to scale). From top to bottom: Stone Industrial kapton straw with $<$~5~$\mu$m thick aluminum layer inside, Euclid Spiral Paper Tube Corp. mylar straw with an aluminum layer (unknown thickness) inside, Lamina mylar straw with a 0.03~$\mu$m aluminum layer inside, and Lamina mylar tube with a 12~$\mu$m aluminum layer inside.}
\end{figure}
A picture of the four different straw types is shown in Fig.~\ref{straws}. The straw type of preference is the Lamina-thin one because this straw has the least amount of material (smallest wall thickness) which means a minimal probability of photon conversion and multiple scattering of particles passing through the CDC.

\subsubsection{Radiation dose}
The source used for irradiation was a 3.7$\cdot 10^{6}$~Bq $^{90}$Sr source. 
The source was collimated and a length of about 1~cm of the straw was 
irradiated. The accumulated dose per straw is shown in Table~\ref{doses}, expressed in C/cm, which is the charge seen by the wire per cm.   
\begin{table}[h]
\begin{center}
\begin{tabular}{|c|c|}
\hline
{\bf Straw type} & {\bf Dose [C/cm]}\\
\hline
Stone-kapton & 0.009  \\
Euclid-mylar & 0.004 \\
Lamina-thin  & 0.010  \\
Lamina-thick & 0.004 \\
\hline
\end{tabular}
\caption{Received doses of the irradiated straws.}
\label{doses}
\end{center}
\end{table}
All the doses are well below 1~C/cm which can rule out traditional 
ageing effects (like the Malter effect) on the straws. These doses are equivalent to about 
44 years of running in the GlueX experiment for the most heavily exposed straws. During irradiation the gas mixture flowing through the prototype was an argon-CO2 (90\%-10\% by volume) mixture and the wire was set at the nominal operating voltage of 1450~V. The straws were grounded. Irradiated straws were later compared to unused (reference) straws that were neither irradiated nor used in a chamber with gas or High Voltage (HV). In the following section the results are discussed. In general, all straws were still operational after irradiation with no performance loss.  More details concerning this test can be found in Ref.~\cite{radtest}.

\subsubsection{Lamina-thin}
\begin{figure}
\begin{tabular}{cc}
\includegraphics[width=0.75\linewidth]{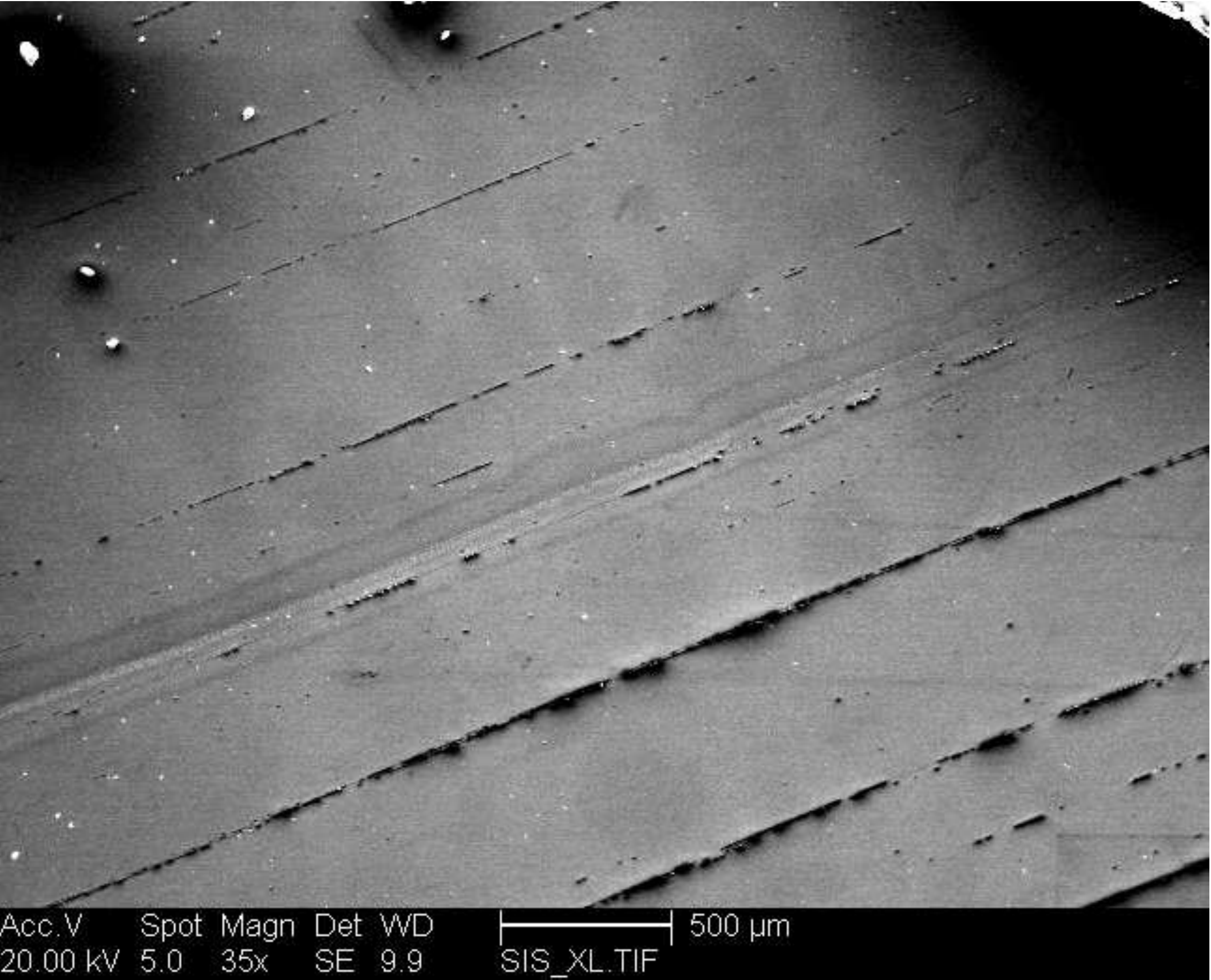}\\
\includegraphics[width=0.75\linewidth]{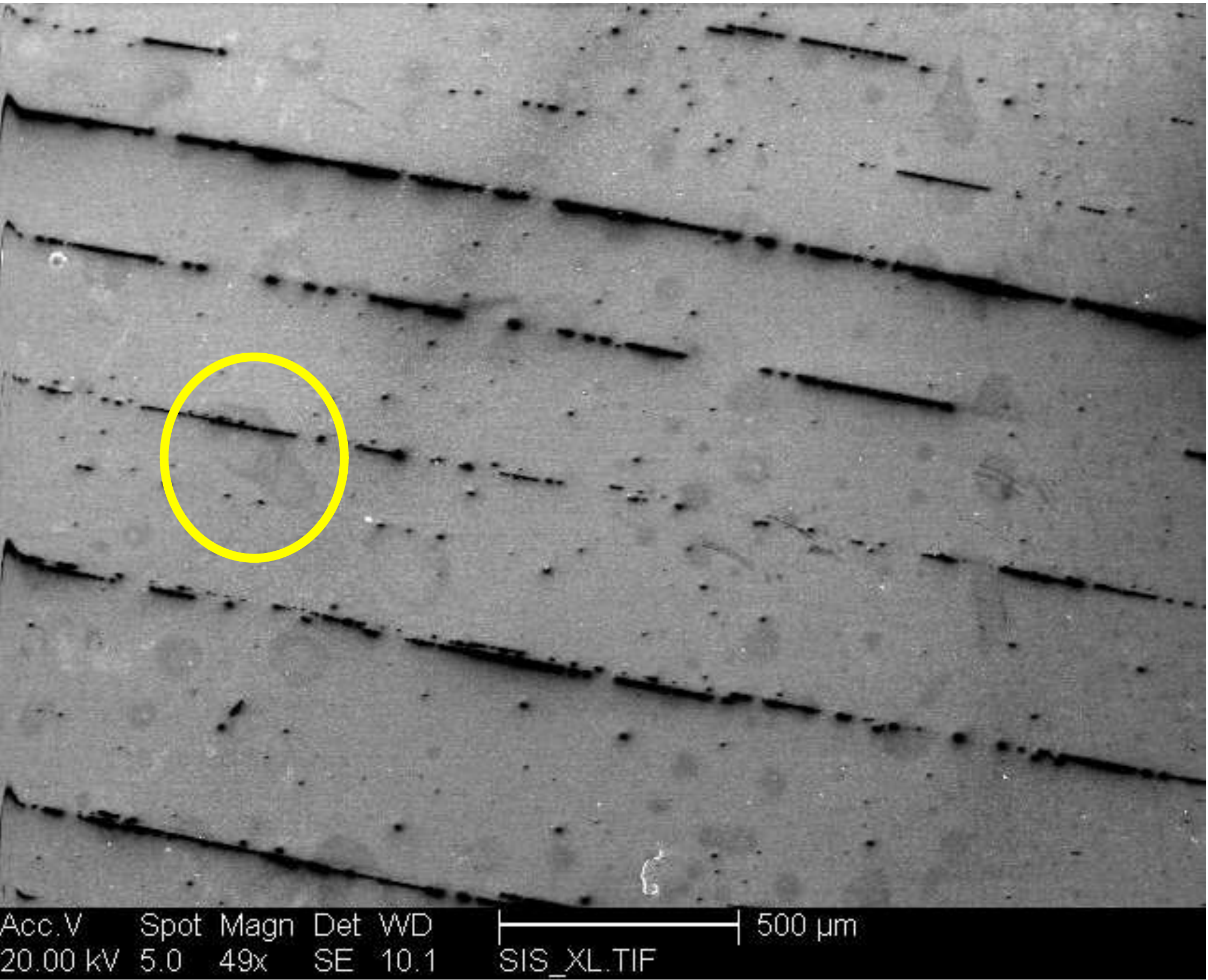}
\end{tabular}
\caption[]{\label{500um} Comparison of the aluminum surfaces between a pristine Lamina-thin straw (top) and one that was irradiated (bottom). The scale of 500~$\mu$m is indicated at the bottom of each picture. A dark gray area, with a black ``line'' going through it, is emphasized by the yellow ellipse.}
\end{figure}

\begin{figure}
\begin{tabular}{cc}
\includegraphics[width=0.75\linewidth]{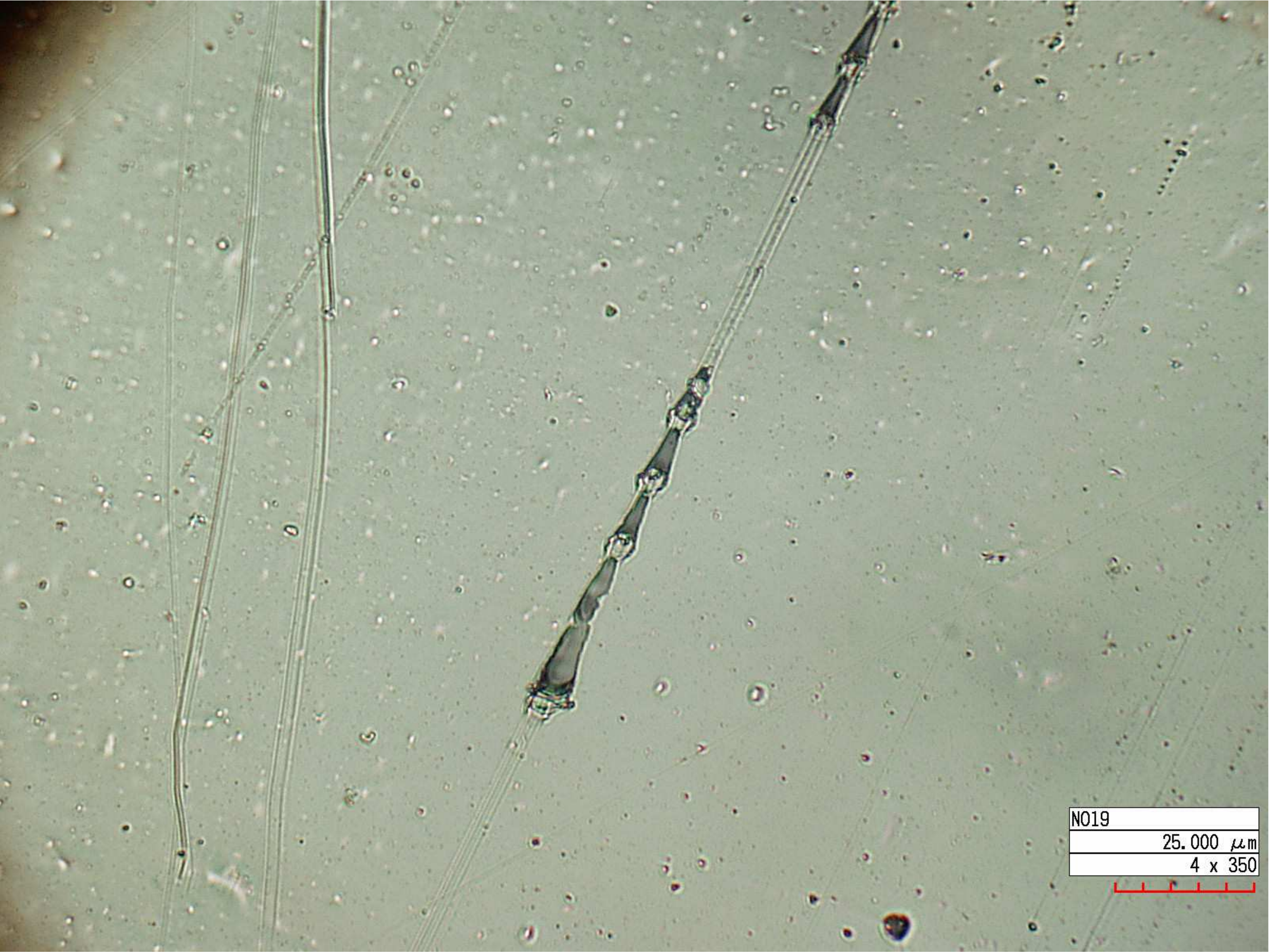}\\
\includegraphics[width=0.75\linewidth]{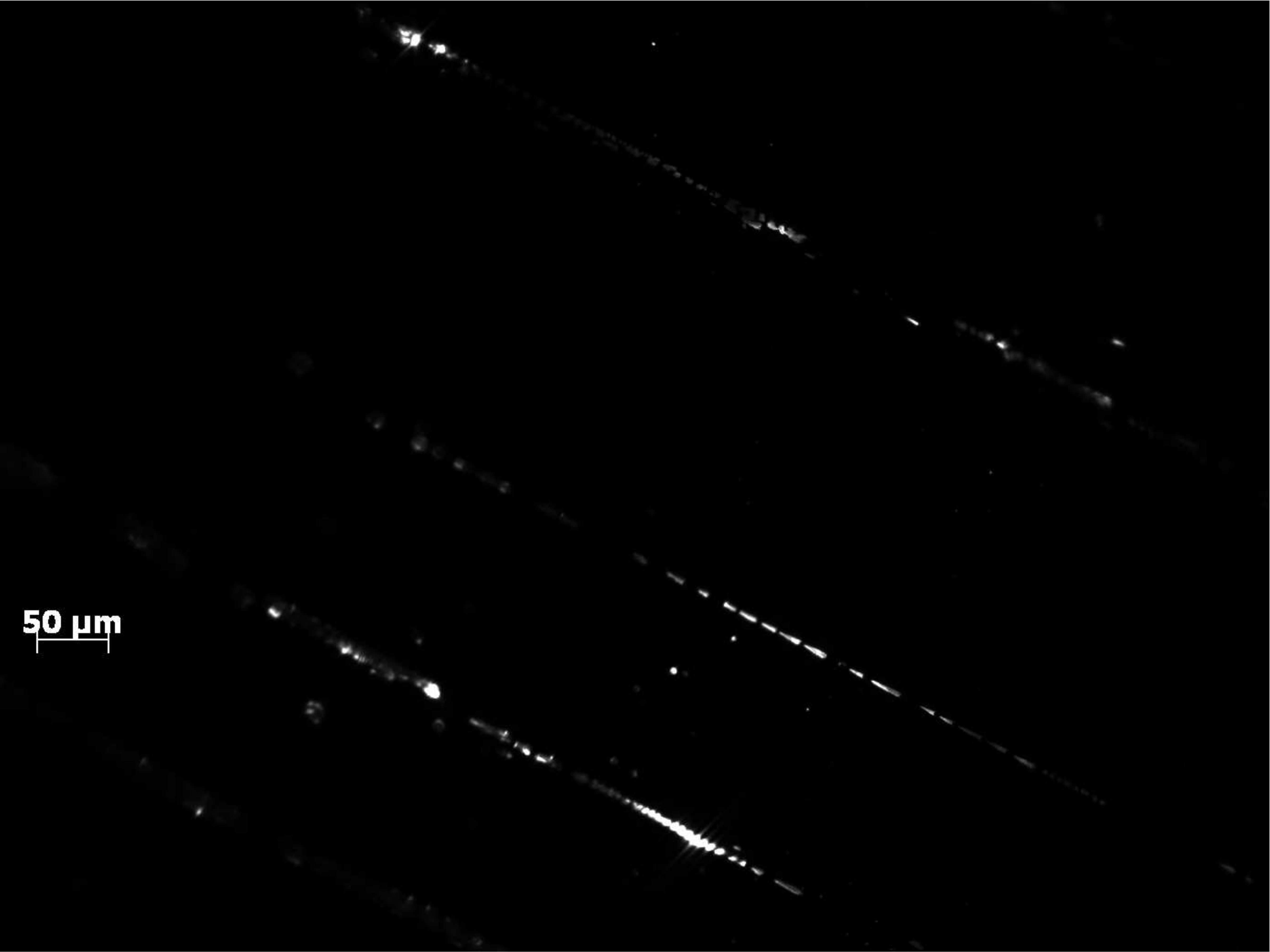}\\
\end{tabular}
\caption[]{\label{optmic} Picture taken of a pristine Lamina-thin straw with an optical microscope. The upper picture shows the straw sample lighted from the top side and a scratch is visible; the red line in the lower right corner indicates 25~$\mu$m. The lower picture shows the straw sample with a light source below it, the white lines and dots are holes or scratches through the aluminum layer.}
\end{figure}

An unused reference Lamina-thin straw was compared, using an electron microscope, with a piece that has been irradiated. Pictures at low scale are shown in Fig.~\ref{500um}.
At this scale black ``lines'' can be seen on both sample surfaces. It is believed that they are an artifact of manufacturing. Pictures of a pristine straw taken with an optical microscope are shown in Fig.~\ref{optmic}. One can see that the black ``lines'' mentioned before are in fact scratches partially or completely through the aluminum layer.    

\begin{figure}
\begin{tabular}{cc}
\includegraphics[width=0.75\linewidth]{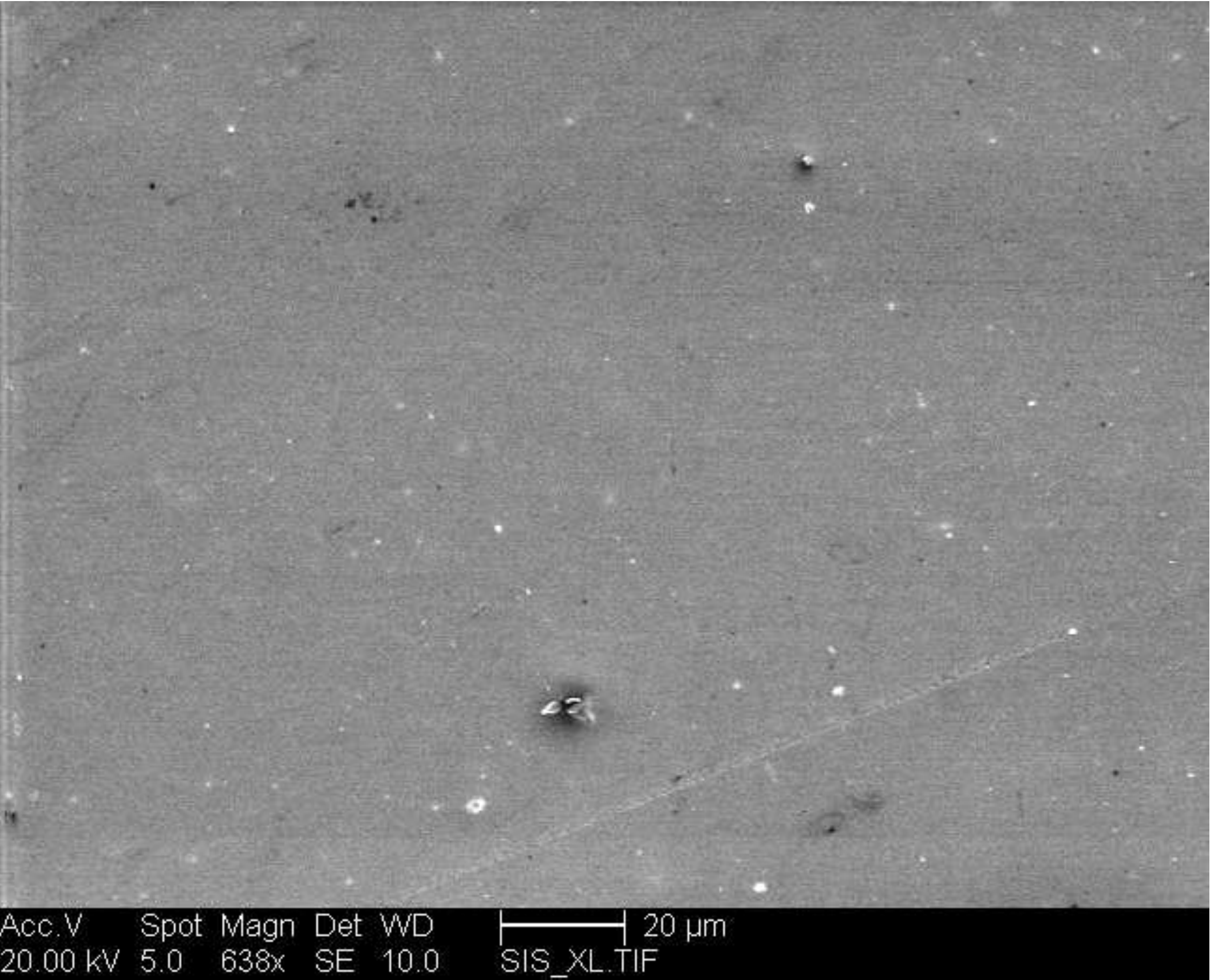}\\
\includegraphics[width=0.75\linewidth]{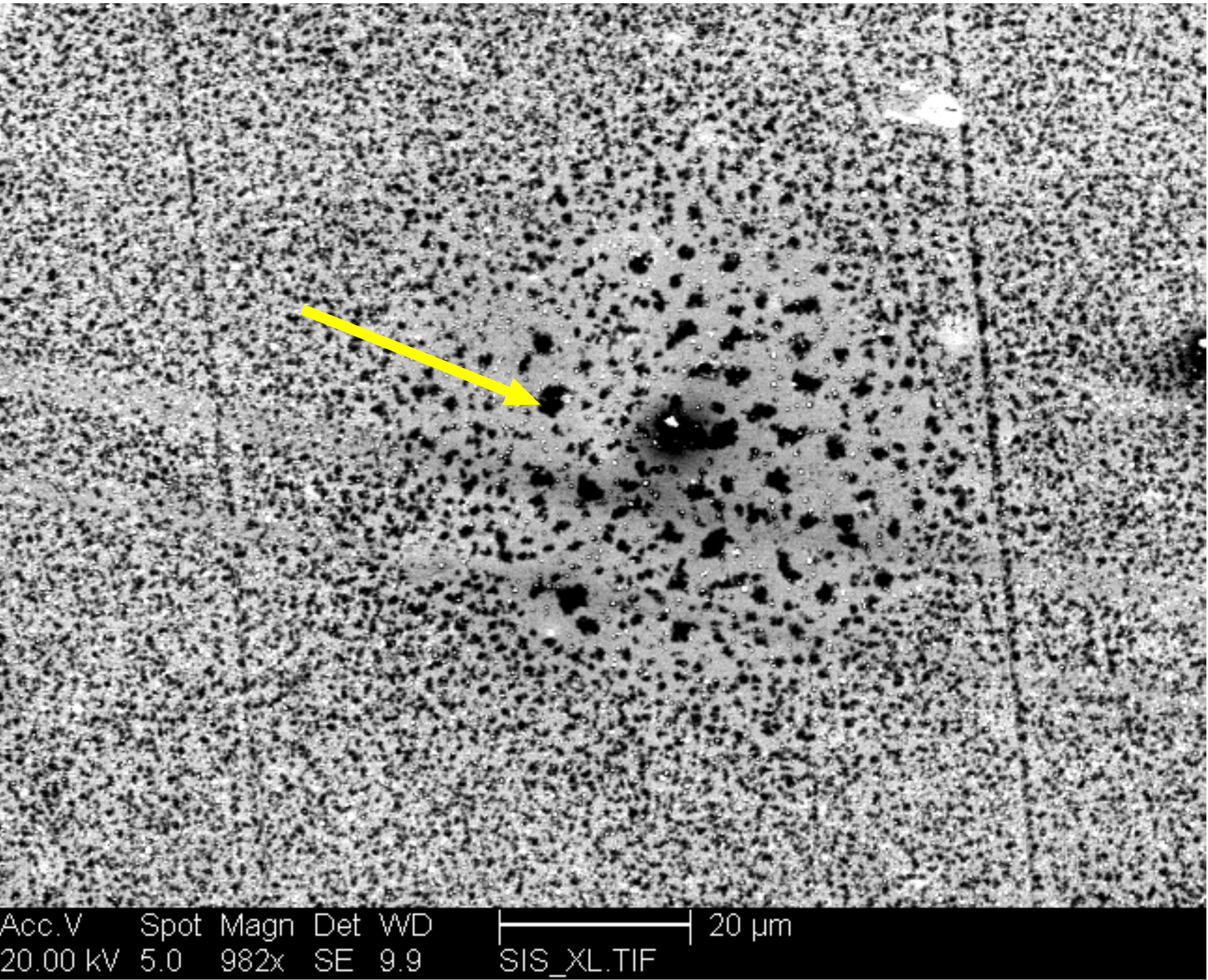}\\
\end{tabular}
\caption[]{\label{5-20um} Comparison of the aluminum surfaces between a pristine Lamina-thin straw (top) and one that was irradiated (bottom) at a scale of 20~$\mu$m, as indicated at the bottom of the pictures. The center of the picture on the bottom is a further magnification of one of the dark gray areas as seen in Fig.~\ref{500um} (bottom). The yellow arrow points to a typical black pit.}
\end{figure}

A difference observed between the pristine and irradiated straw are the dark gray areas (one area is indicated with a yellow ellipse in Fig.~\ref{500um}) on the irradiated straw which are not visible on the reference straw. In Fig.~\ref{5-20um}, at a factor of 20 higher resolution, one can see that the irradiated straw shows 
black pits and that the dark gray areas indicated in Fig.~\ref{500um} are in fact a different size and concentration of those black pits (as can be seen in the center of Fig.~\ref{5-20um}). The origin of the pits remains unclear but they have a relative higher amount of oxygen present than the surrounding area. The relative amount of atoms can be determined by doing an analysis of X-rays coming from the straw when under the electron microscope, three lines with different intensity are always observed: aluminum, oxygen, and carbon. The pits are also observed in straw parts that were used in the radiation exposure test but were located far away (20~$cm$) from the Sr-source and in a Stone-kapton straw that was used for three years but then the effect was less intense and no difference in oxygen content could be measured. This suggests that the spots are caused by a chemical reaction triggered by radiation and the presence of an electric field.

\subsubsection{Lamina-thick}
\begin{figure}
\begin{tabular}{cc}
\includegraphics[width=0.75\linewidth]{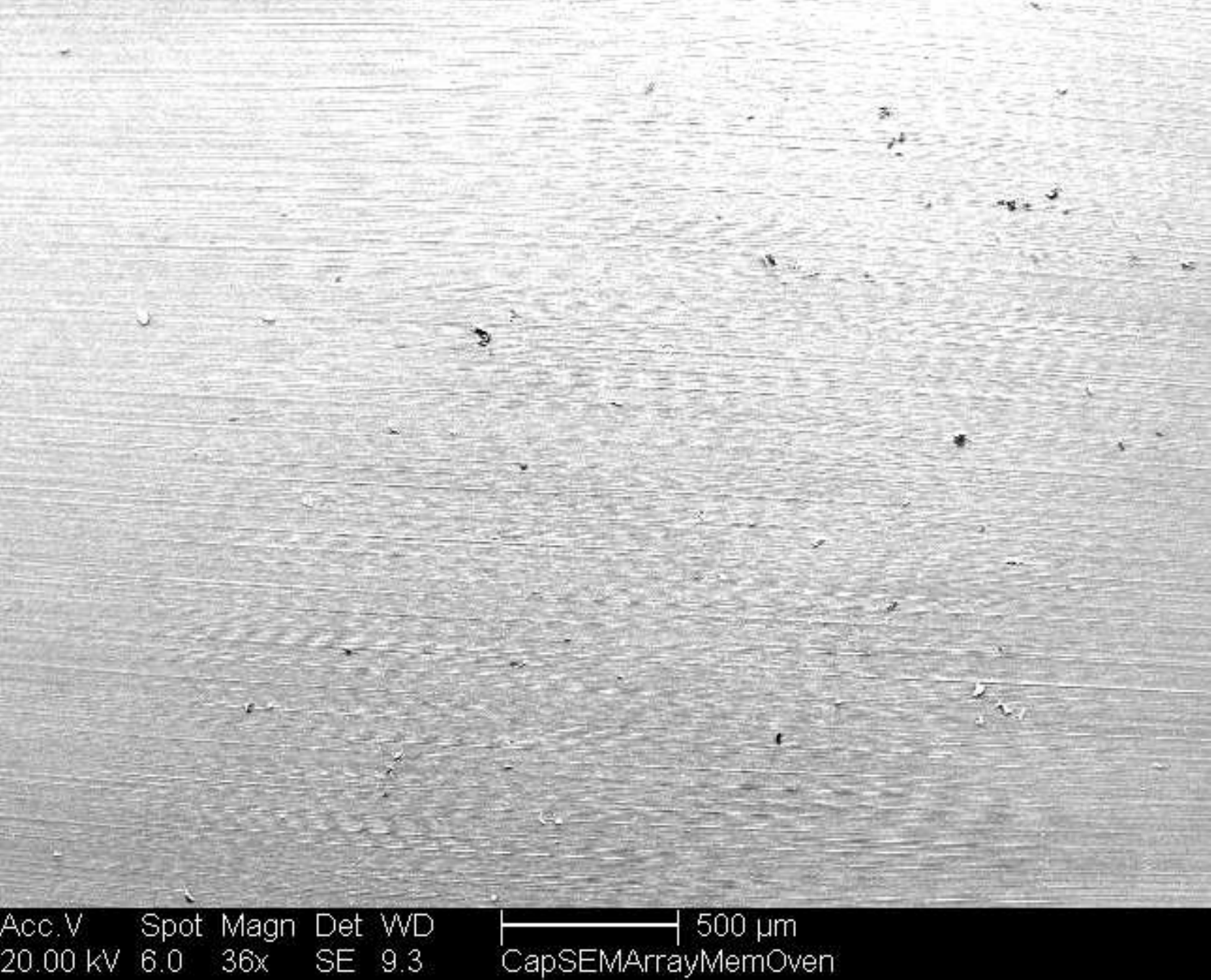}\\
\includegraphics[width=0.75\linewidth]{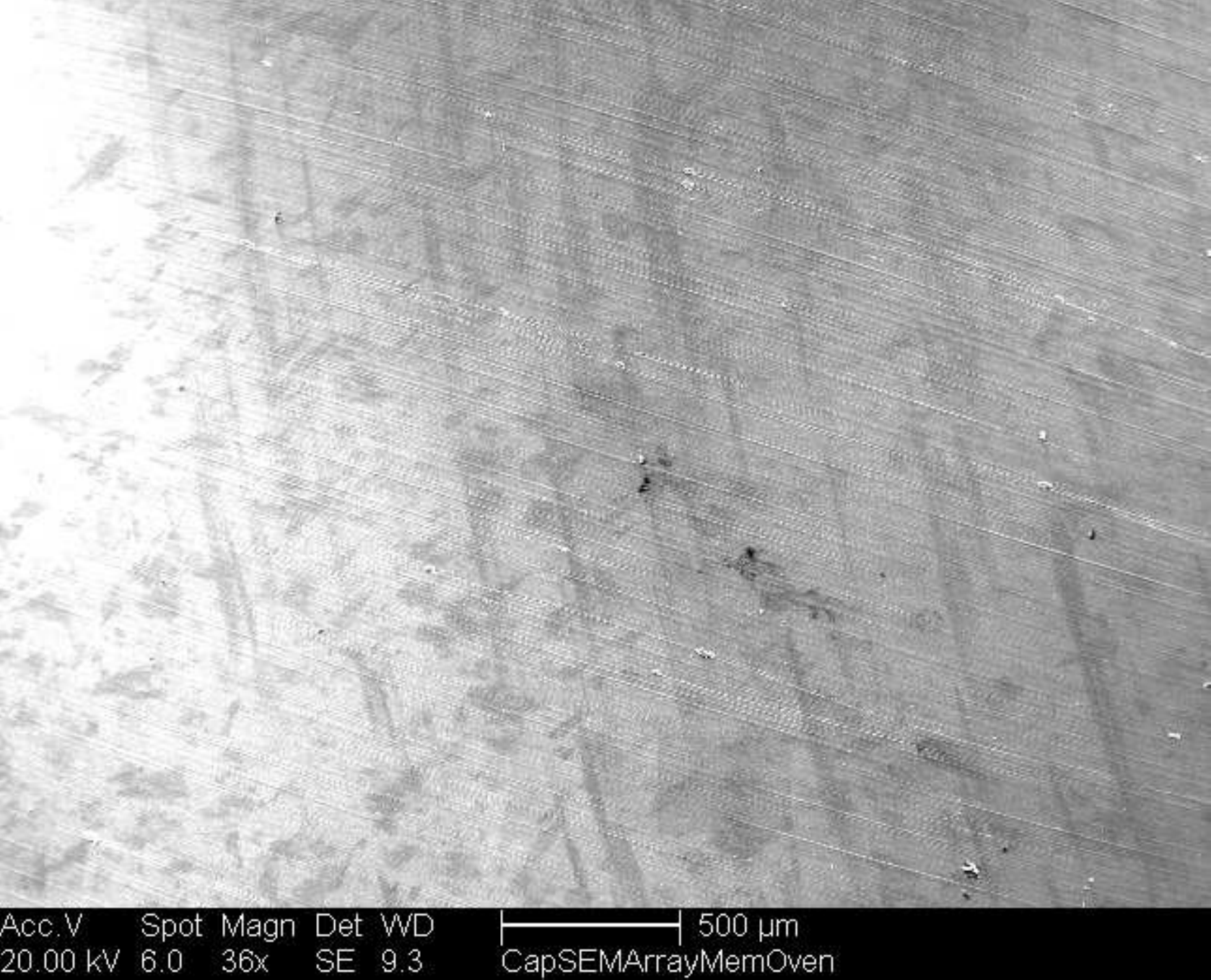}
\end{tabular}
\caption[]{\label{500um-thin} Comparison of the aluminum surfaces between a pristine Lamina-thick straw (top) and the one that was irradiated (bottom) at a scale of 500~$\mu$m as indicated at the bottom of the pictures.}
\end{figure}

\begin{figure}
\begin{tabular}{cc}
\includegraphics[width=0.75\linewidth]{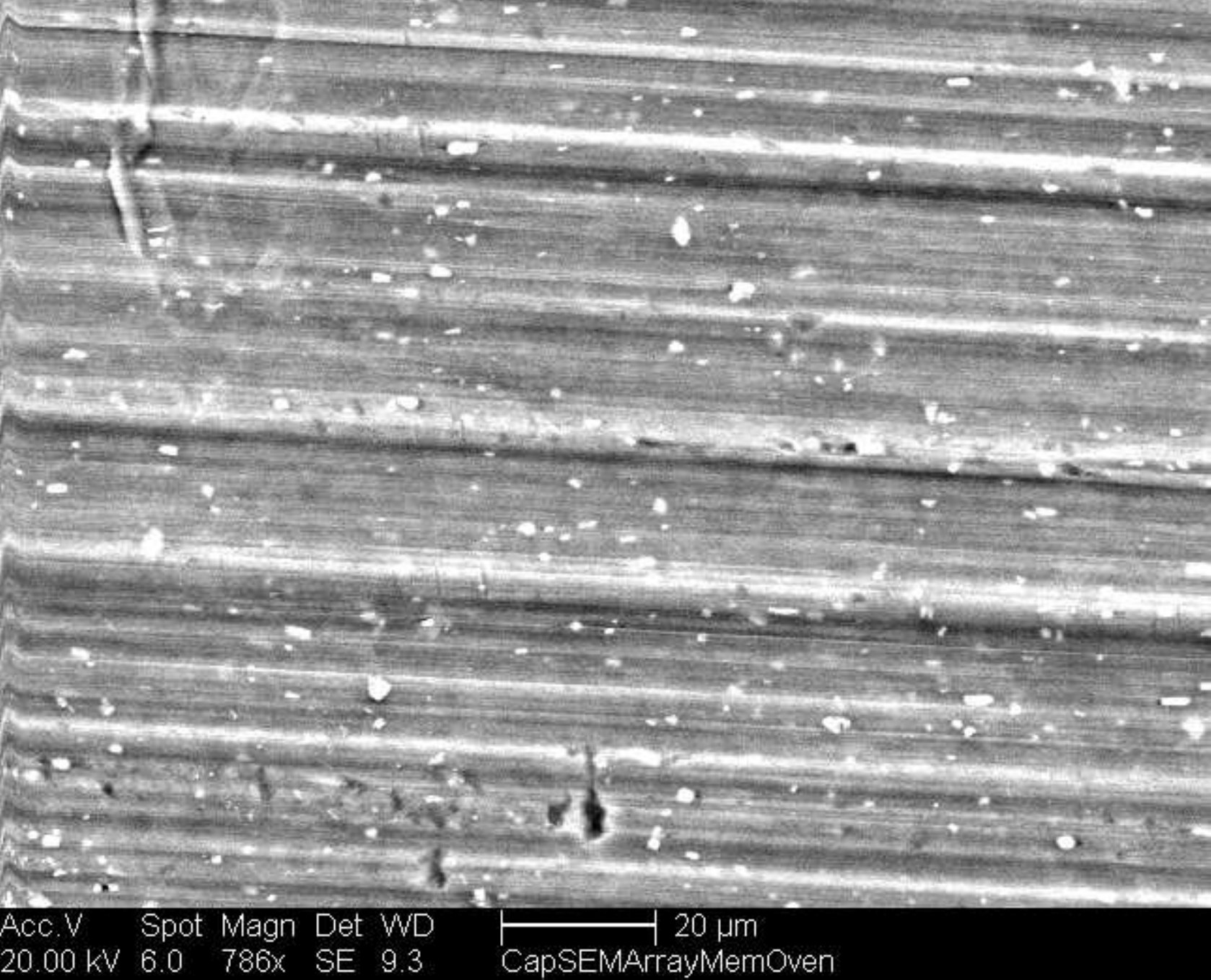}\\
\includegraphics[width=0.75\linewidth]{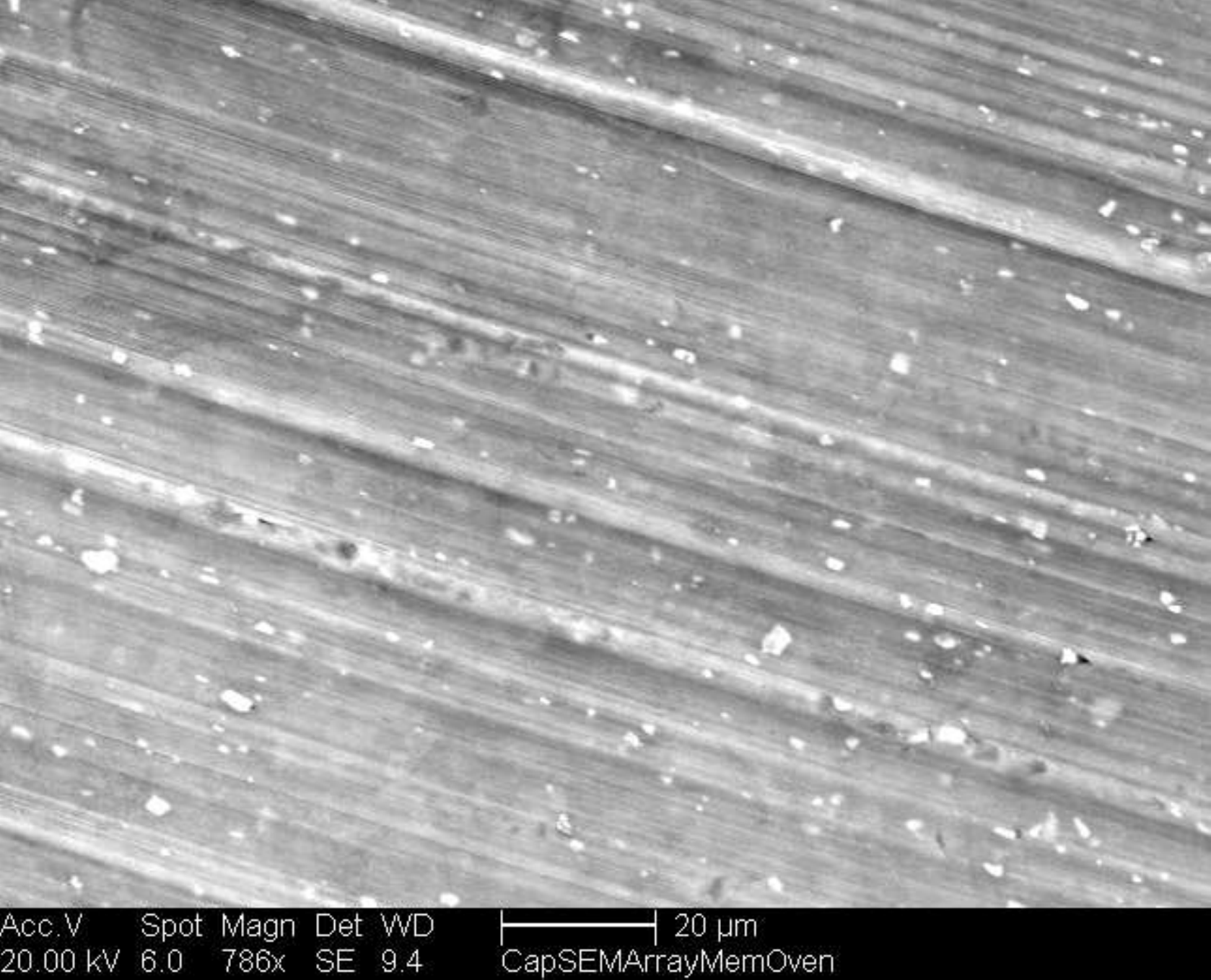}\\
\end{tabular}
\caption[]{\label{5-20um-thin} Comparison of the aluminum surfaces between a pristine Lamina-thick straw (top) and the one that was irradiated (bottom) at a scale of 20~$\mu$m as indicated at the bottom of the pictures.}
\end{figure}

A piece of a pristine Lamina-thick straw is compared, using an electron microscope, to a piece that was irradiated. The result at low resolution is shown in Fig.~\ref{500um-thin}. Both straw surfaces look very smooth compared to the same pictures taken of the 
Lamina-thin straw.
On this scale one can see that there is no substantial difference between the two. Also at higher magnification no differences can be seen (Fig. \ref{5-20um-thin}).  
The white dots are little pieces of pure iron, probably stemming from fabrication. The difference with the previous straw (Lamina-thin) is, according to the manufacturer, that there is an aluminum foil being glued to a mylar sheet instead of being vapor-deposited. There were no holes or scratches observed that go through the aluminum layer when looking at it with an optical microscope with a light source mounted underneath the straw.

\subsubsection{Gold plated tungsten wire}
\begin{figure}
\begin{tabular}{cc}
\includegraphics[width=0.75\linewidth]{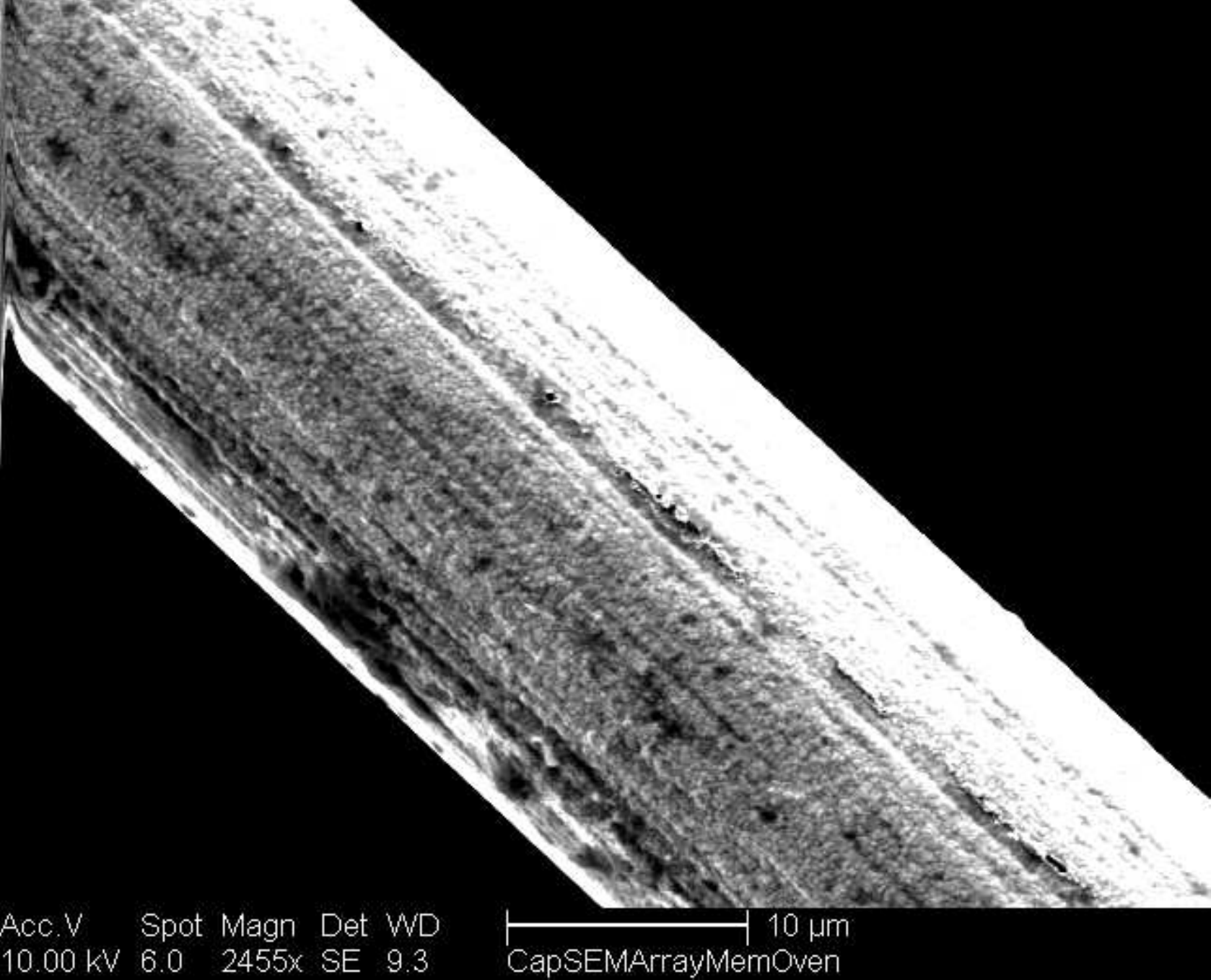}\\
\includegraphics[width=0.75\linewidth]{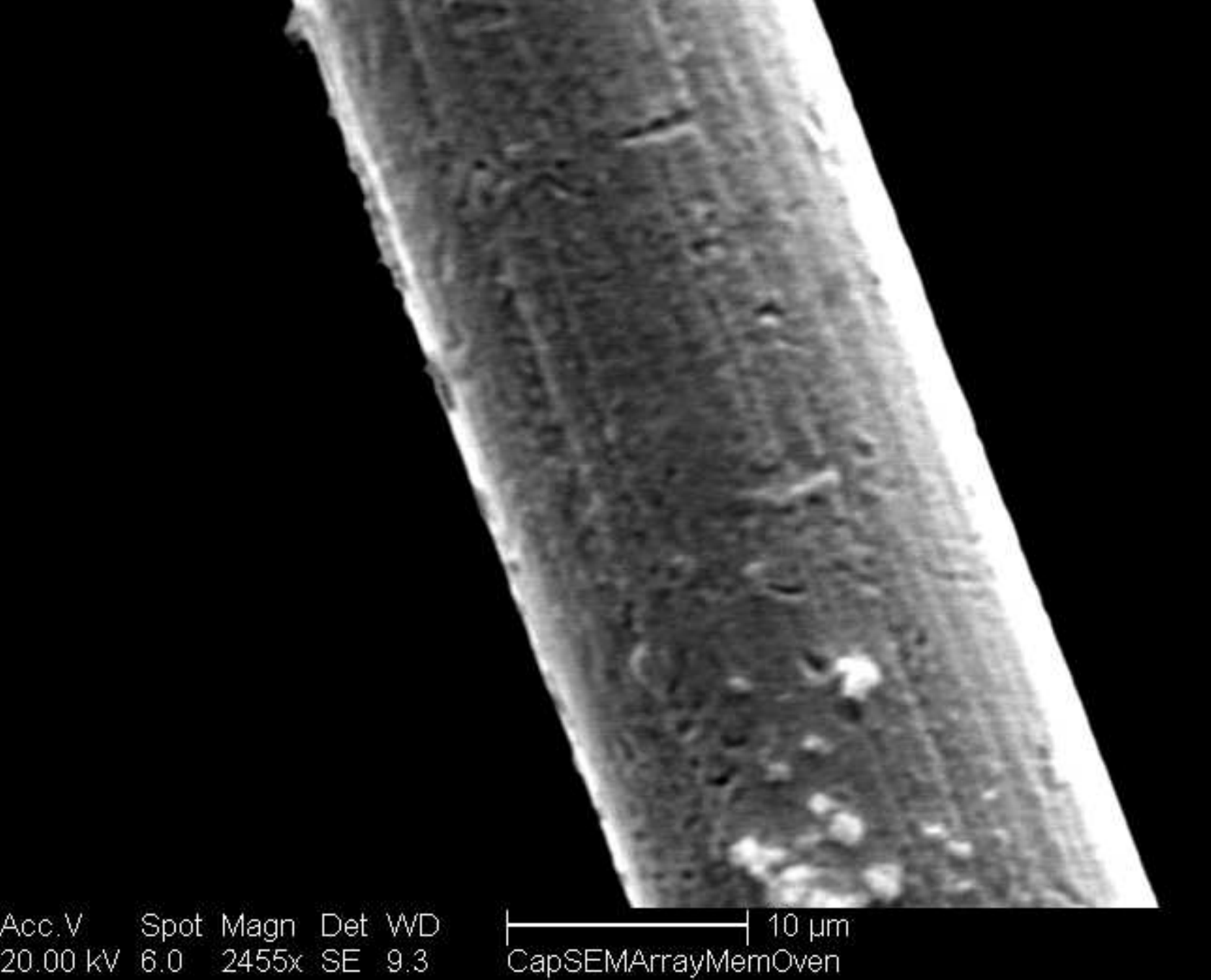}\\
\end{tabular}
\caption[]{\label{wirecomp} Comparison between a pristine wire (top) and one that was irradiated (bottom) at a scale of 10~$\mu$m as indicated at the bottom of the pictures.}
\end{figure}

The 20~$\mu$m diameter gold plated tungsten wire was also examined before and after the radiation test. A comparison is shown in Fig.~\ref{wirecomp}. As the radiation doses are low (much lower than 1~C/cm) no radiation damage is to be expected. Indeed there is no obvious difference observed between the irradiated and the pristine wire.

\subsection{Straw choice}
The straw type choice is not straightforward as the Lamina-thin one has the least amount of material but shows scratches, holes and some unexpected behavior after radiation. However, the straw was still operational after radiation with no performance loss. The tests performed do not give any information of how this will evolve over time. The Lamina-thick straw shows a very stable behavior but the aluminum layer is too thick. The straw of choice will be the Lamina-thin one but with a thicker vapor-deposited aluminum layer, probably between 0.5 and 1~$\mu$m. They will be checked before usage. The total wall thickness of this straw will be between 100 and 101~$\mu$m depending on the thickness of the inner aluminum layer. The inner radius of the straw is 7.8~mm.

\subsection{Straw placement}
\begin{figure}
\begin{tabular}{cc}
\includegraphics[height=1.\linewidth, width=0.9\linewidth]{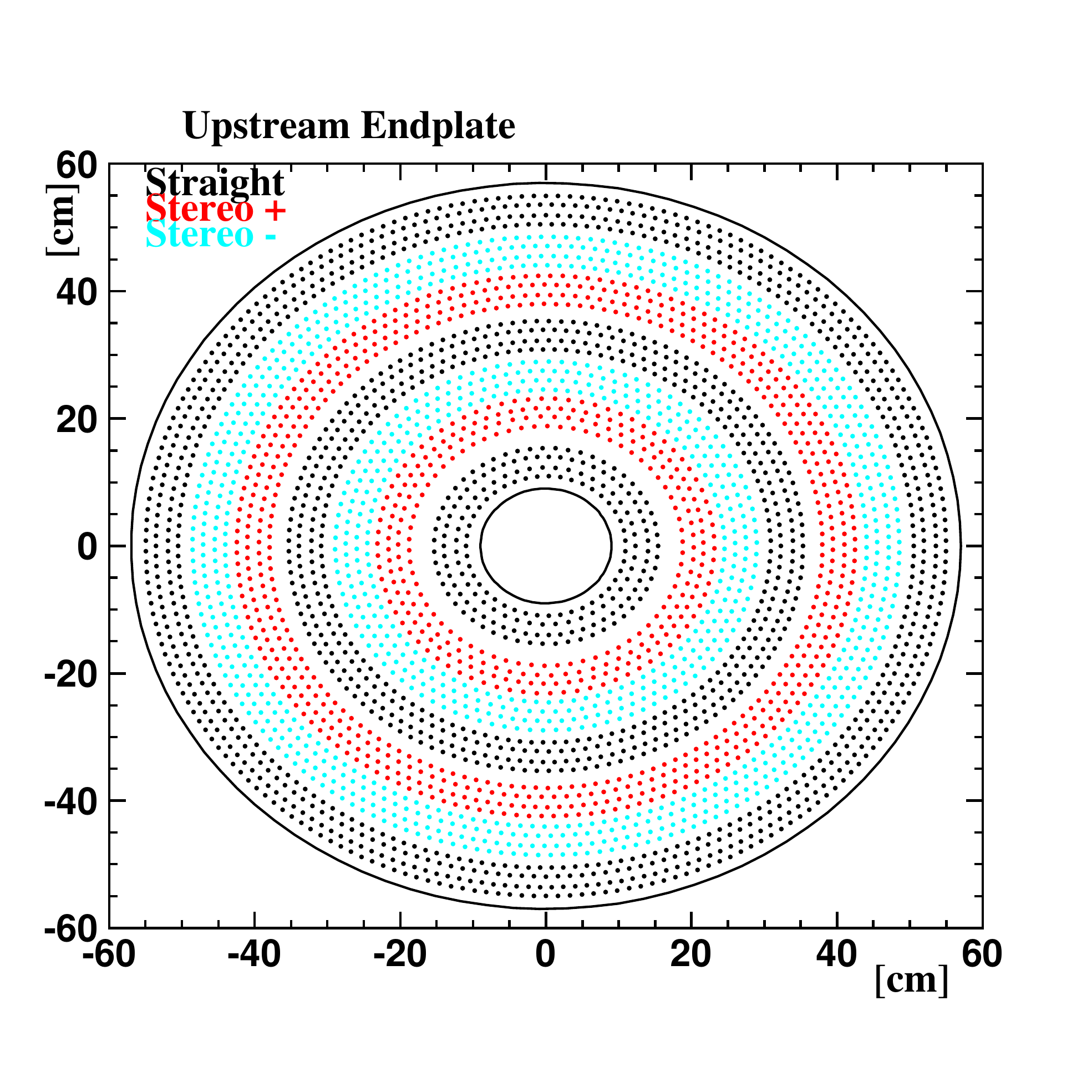}
\end{tabular}
\caption[]{\label{strawplac} Drawing of the front view of straw position in the upstream end plate. Black dots represent axial straws, red and blue dots represent straws with a stereo angle of plus and minus 6$^{\circ}$ respectively. } 
\end{figure}

The straws will be arranged in 28 layers. From the inside out: two double (a layer followed by a close packed layer) axial layers, followed by two sets of: {two double +6$^{\circ}$ stereo layers, two double -6$^{\circ}$ degree layers and two double axial layers}. A drawing showing the placement of the straws is shown in Fig.~\ref{strawplac}. This setup has been carefully chosen after extensive Monte Carlo (MC) simulations in order to minimize tracking ambiguities \cite{mclayer}. The innermost layer of straws is at a radius of 9~cm and the outermost layer at a radius of 59.75~cm.  
\section{Gas system and flow}
The gas mixture that is intended to be used is a mixture of argon and CO$_2$. Flammable gas components are avoided for safety reasons. The gas system to flow the gas mixture through the prototypes consists of two MKS Instruments\footnote{MKS instruments, 2 Tech Drive, Suite 201, Andover, MA 01810, http://www.mksinst.com} Mass-Flo controllers one for argon and one for CO$_2$ gas. They are read-out and controlled by a MKS Instruments Inc. Multigas controller. The gas is mixed in a buffer volume of 34.4~l before being fed into the downstream end plenum of the detector. On the upstream and downstream end of the CDC a gas plenum is constructed where all the straws end. This construction allows to have one gas inlet into the downstream gas plenum to feed gas to all the straws and one gas outlet on the upstream end plenum to extract the gas from all the straws. In the GlueX experiment the gas flows as follows: gas will flow into the downstream end plenum through 6 plastic tubes. Gas then flows through the straws into the upstream end plenum where it can flow through 6 holes in the upstream end plate into the volume between straws. This volume will be made gas tight. Gas is extracted from this volume though 3 plastic tubes going through the upstream end plenum and end plate. The pressure is slightly above 1~atm and the gas flow in the prototype is a few detector volumes per hour. About three volume exchanges per day are anticipated for the gas system in the GlueX detector~\cite{gasflow}. The gas temperature and pressure will be monitored.  
\section{Electronics}
The inside aluminum surfaces of the straw tubes are connected to ground while the wires are held at a voltage between 1400 and 2300~V (depending on the gas mixture). The wires are connected to cables that run through the end plenum to HV-boards~\cite{hvboard} located outside of the gas volume. These printed circuit boards distribute high voltage to the wires and decouple the signal from the high voltage line (it also protects against high currents and filters high frequency noise on the high voltage line). This board hosts an ASIC charge-sensitive pre-amplifier~\cite{asic}. The charge-amplified signal of the current prototypes is shaped by a shaper board~\cite{shaper} and readout by Struck\footnote{Struck Innovative Systeme GmbH, Harksheider Str. 102A, 22399 Hamburg, Germany, http://www.struck.de} SIS3320 12 bit flash ADCs running with a 125~MHz external clock. They are run at 125~MHz (they can run up to 200~MHz) because in the GlueX experiment the shaper and the flash ADC will be combined into one VME module~\cite{fadc} that will run with an 125~MHz external clock. This module is being designed and tested at Jefferson Lab.    
\section{Prototypes}
\begin{figure}
\begin{tabular}{cc}
  \includegraphics[width=0.95\linewidth]{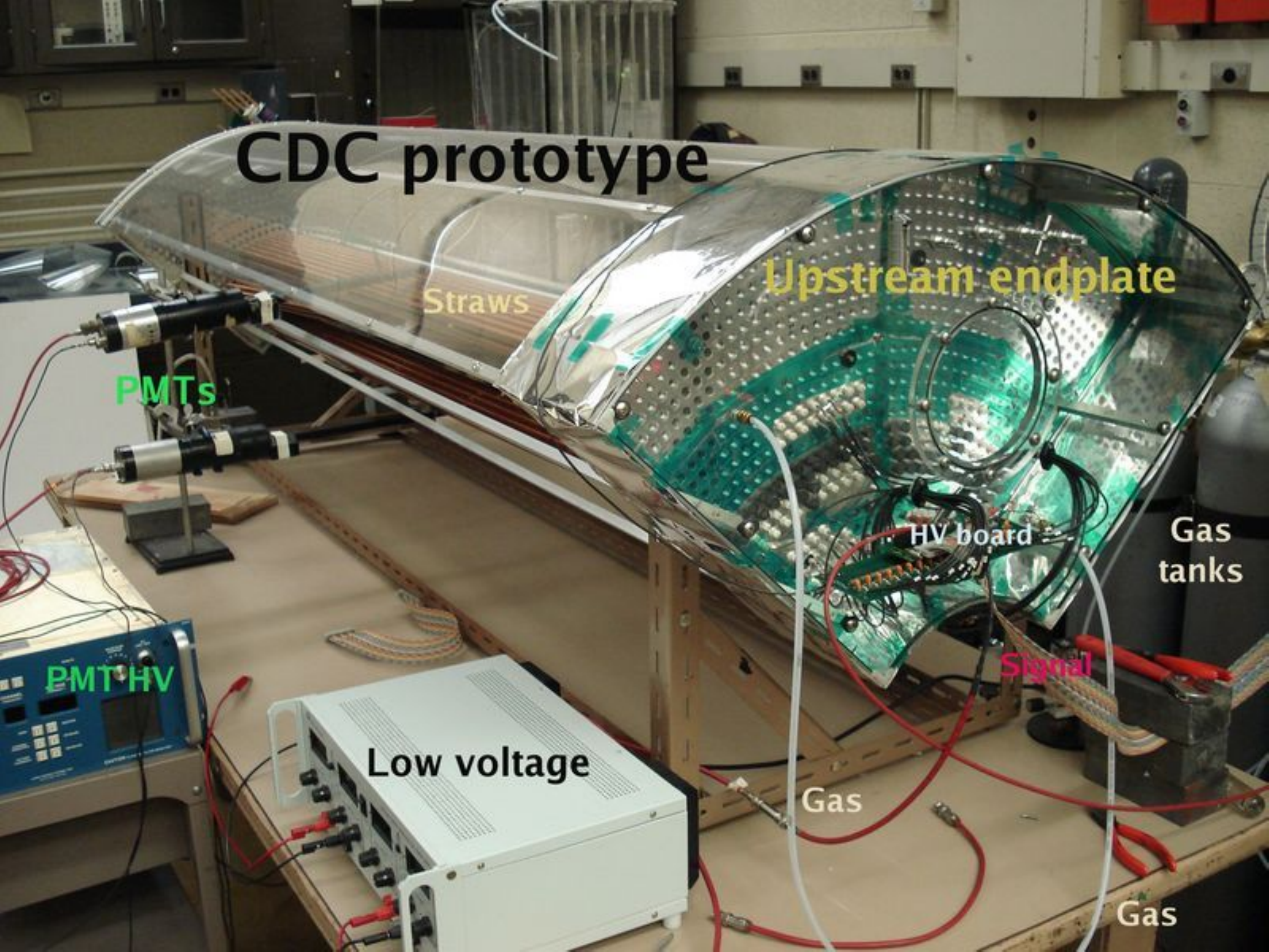}
\end{tabular}
\caption[]{\label{fullprot} Picture showing a one quarter section of a full-length prototype of the CDC.}
\end{figure}

Three different prototypes have been constructed. First, a full-length prototype of one
quarter of the final CDC was built with two different straw types: Stone-kapton and Euclid-mylar straws. The straws are 2~m long (in the final design the straws will be 1.5~m long). A picture of this prototype is shown in Fig.~\ref{fullprot}. This prototype was mainly constructed to confirm the feasibility of the design and to optimize details in the construction of the detector. It turned out that the Euclid-mylar straws can be easily deformed and the Stone-kapton straws not. Tests were also done to compare the options of placing the electronics either inside or outside the gas plenum. Both configurations could be implemented in the experiment with acceptable distance between the wire and the pre-amplifier so it was decided to locate the electronics outside the gas plenum to facilitate access. Operating this prototype revealed the necessity for generous solid ground connection between the end-plate that connects to the aluminum surfaces of the straws and the pre-amplifier to minimize electronic noise.

\begin{figure}
\begin{tabular}{cc}
  \includegraphics[width=0.85\linewidth]{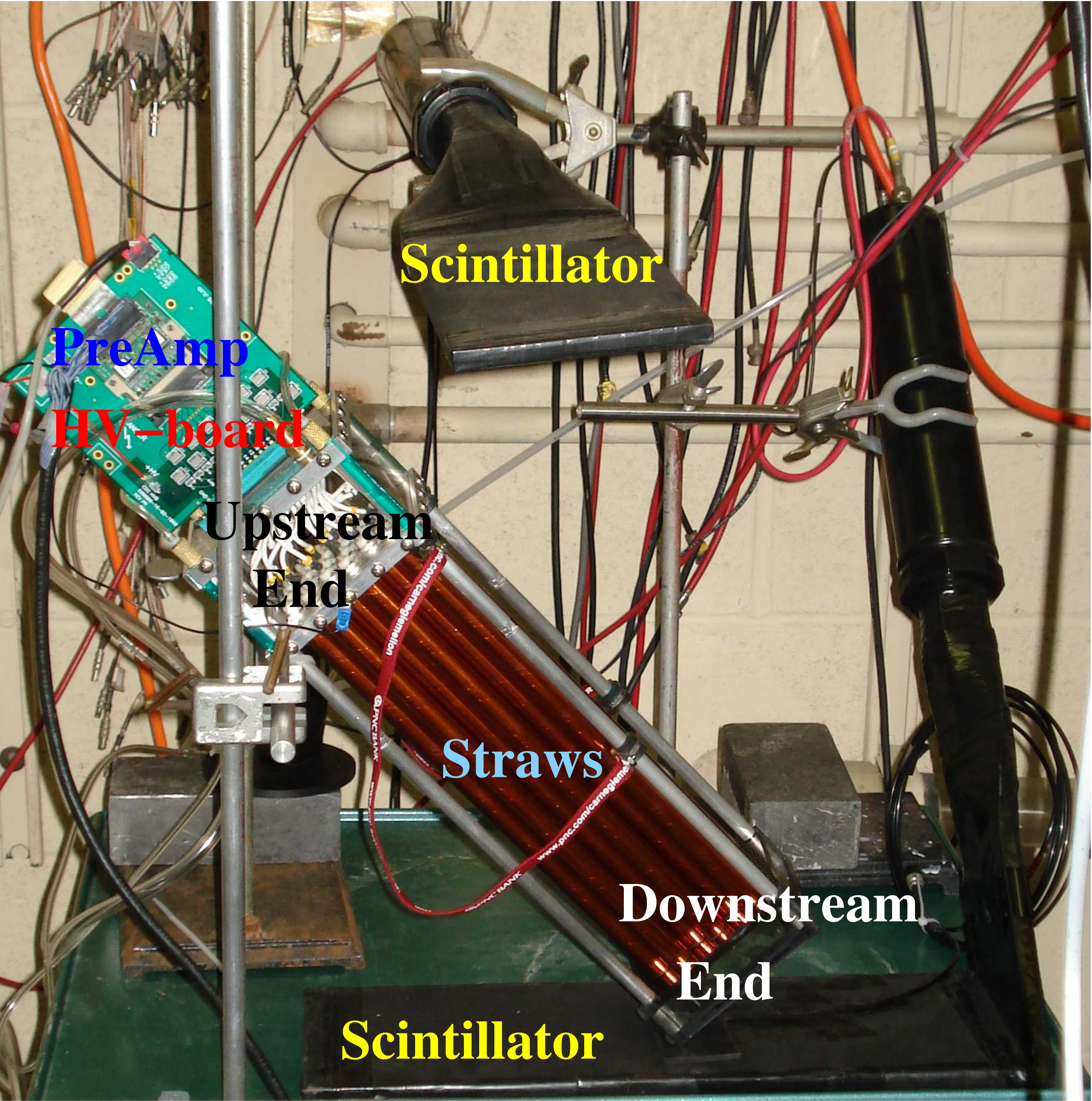}\\
  \includegraphics[width=0.85\linewidth]{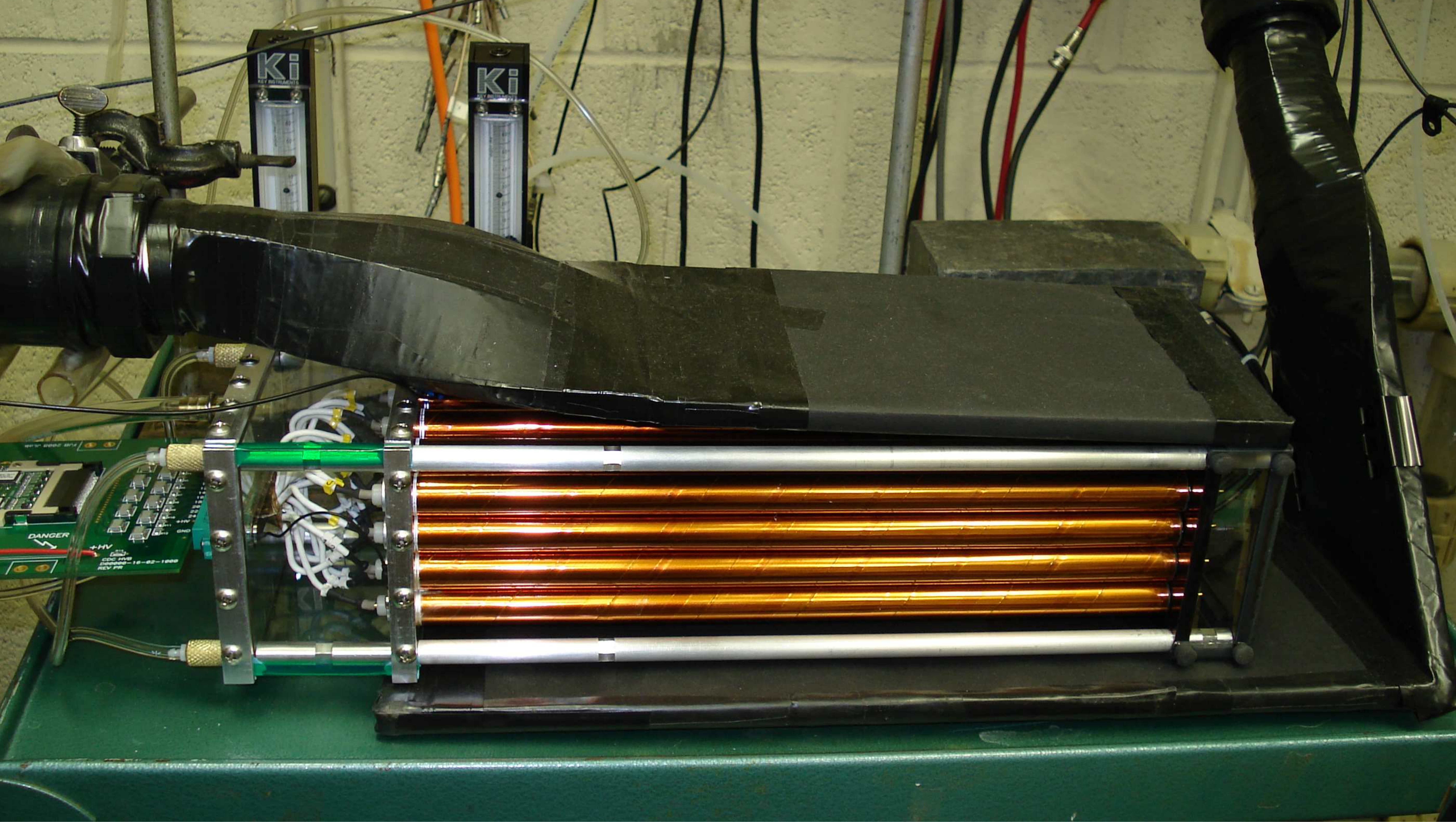}
\end{tabular}
\caption[]{\label{smallprot1} Pictures showing a small scale prototype of the CDC with a scintillator above and below (for cosmic ray triggering) the prototype. Above the prototype is tilted and the different parts are labeled, at the bottom side the prototype is positioned horizontal with a scintillator mounted above and below the chamber.}
\end{figure}

\begin{figure}
\begin{tabular}{cc}
  \includegraphics[width=0.75\linewidth]{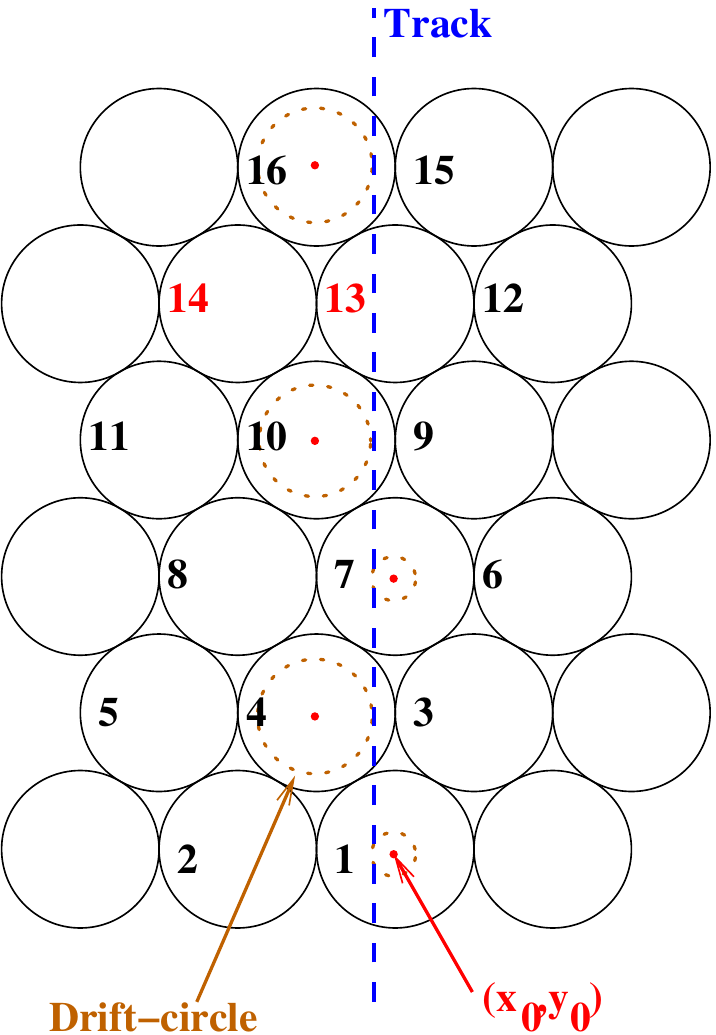}
\end{tabular}
\caption[]{\label{smallprot2} View of the cross section of the close packed straw arrangement of the small scale prototype. The channels 13 and 14 indicate broken channels in the readout electronics while unnumbered straws are not connected to any readout electronics. A possible track and drift-circles illustrate the tracking mechanism in the $x$-$y$ plane, see text for more details.}
\end{figure}

Design problems identified in the full-length prototype are corrected and improvements are implemented in the small prototype (shown in Fig.~\ref{smallprot1}). The small prototype has 24 straws that are 30~cm long with two gas end-plenums. The electronics are mounted outside the ``upstream'' gas plenum. The straws are in a close packed configuration as shown in Fig.~\ref{smallprot2} but only 14 of the 24 channels are connected due to limited availability of readout channels. 

\section{Characterizing the detector}
The following sections describe measurements to determine characteristics which must be determined before more complicated studies can be performed. First, the detector response to a $^{55}$Fe source is studied. Then, gas gains are calculated.  
\subsection{$^{55}$Fe energy spectra}
\begin{figure}
\begin{tabular}{cc}
  \includegraphics[width=1\linewidth]{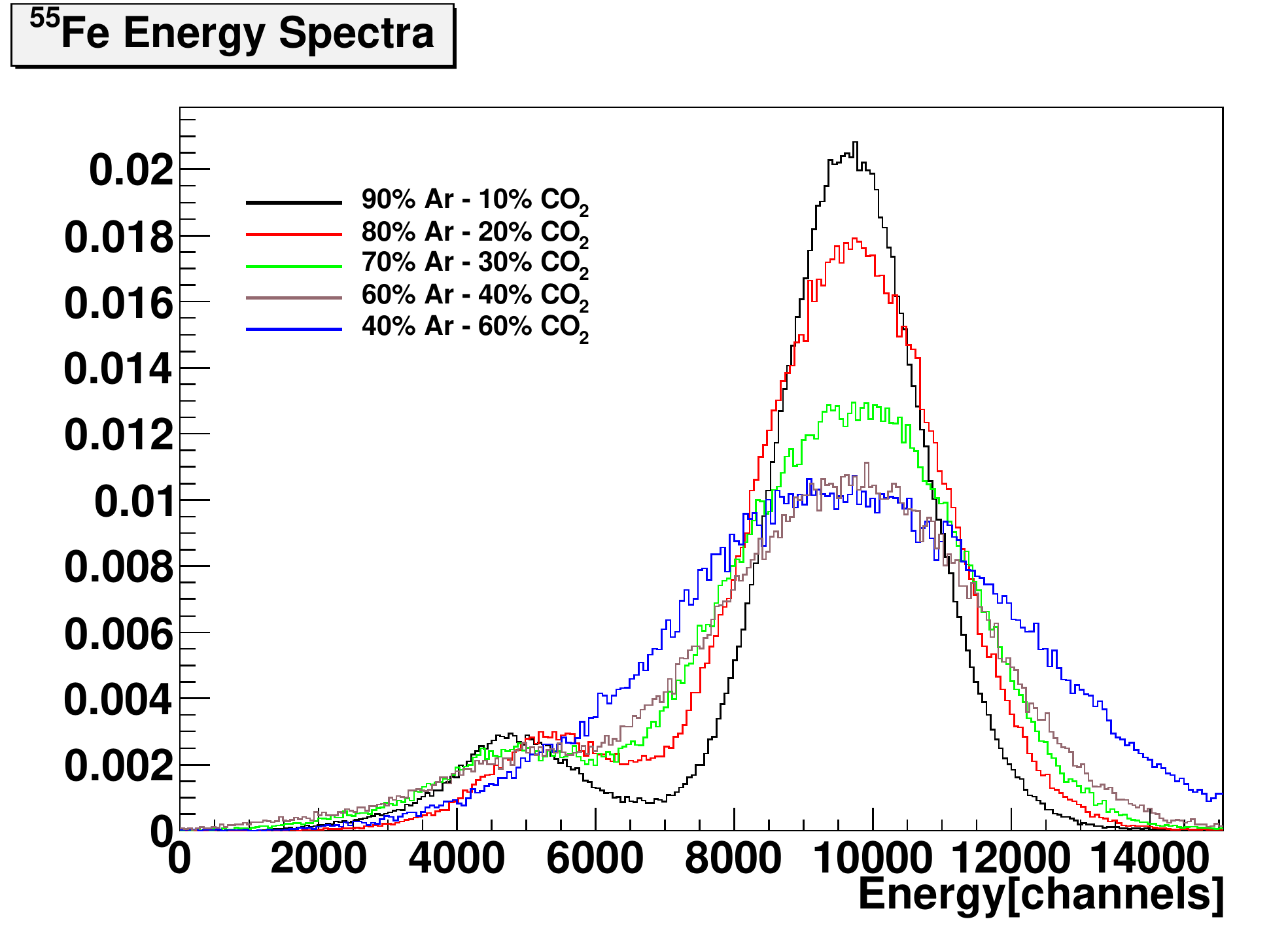}
\end{tabular}
\caption[]{\label{fe55spect} Detector response to the $^{55}$Fe source recorded by the flash ADC for different argon-CO$_2$ gas mixtures. The spectra are corrected for slight gas gain differences and are normalized to each other.}
\end{figure}
Fig.~\ref{fe55spect} shows the detector response to a $^{55}$Fe source using several different argon-CO$_2$ gas mixtures. One can identify a large peak (the argon-peak) at the high end of the energy spectrum. This is the 5.966~keV peak where the X-ray (from the $^{55}$Fe source) is being absorbed by the argon atom and two electrons are freed. The second peak, at lower energy, is the argon-escape peak (the excited argon atom emits an X-ray that has a high probability of escape) and has an energy of about 2.9 keV. The energy resolution degrades with increasing CO$_2$ admixture as can be seen by the increasing width of the argon peak. This behavior can be parameterized with a linear function:

\begin{equation}
  \sigma_E = 815 + 27.5\cdot x_{\rm CO_2},
\end{equation}
with $\sigma_{E}$ the width of the argon-peak and $x_{\rm CO_2}$ the amount in \% of CO$_2$ in the argon-CO$_2$ gas mixture.

\subsection{Monte Carlo simulation}
\label{MCS}
The MC simulation of the small scale prototype is written using {\sc Garfield}~\cite{GARF}. In this simulation a track is generated by two points, one above and one below the detector. Those uniformly generated random points are connected with a straight line and it is determined which straws are intersected by a track. The distance from the track to the wire in the straws that are hit is then passed to {\sc Garfield} which returns output currents as a function of time. Those currents are convoluted with a response function of the electronics chain up to the flash ADC. This function is obtained experimentally by injecting a fast step function signal from a function generator into the readout chain and sampled at its output using an oscilloscope. The amplitude of this signal is sampled at a rate of 125~MHz, the same frequency as the flash ADC, and written to disk in the same format as the data acquisition system. In this way the same code can be used to analyze real data as well as simulated data and the two can be directly compared. 
An advantage of this method is that it can include the effects of the magnetic field on the drift chamber response using {\sc Garfield} but still work with straight tracks. {\sc Garfield} will add drift to the electrons' path as if there was a magnetic field present. In this way, straight line tracking can still be used.

\section{Gas Choice}
In the following section, more complex measurements are described that directly influence the choice of the operating gas mixture. First, the working voltage for each gas mixture is determined. This is followed by gas gain, timing, and resolution studies. Scintillators are placed above and below the prototype, as shown in Fig.~\ref{smallprot1}. Typically, one chooses a noble gas (argon) and a quencher gas (CO$_2$). The quencher absorbs photons originating from excited noble gas atoms. In general these photons have an energy large enough to free electrons from the aluminum inside the straw, this can cause sparks and limits the gas gain.

\subsection{Operating voltage}
\begin{figure}
\begin{tabular}{cc}
  \includegraphics[width=1.0\linewidth]{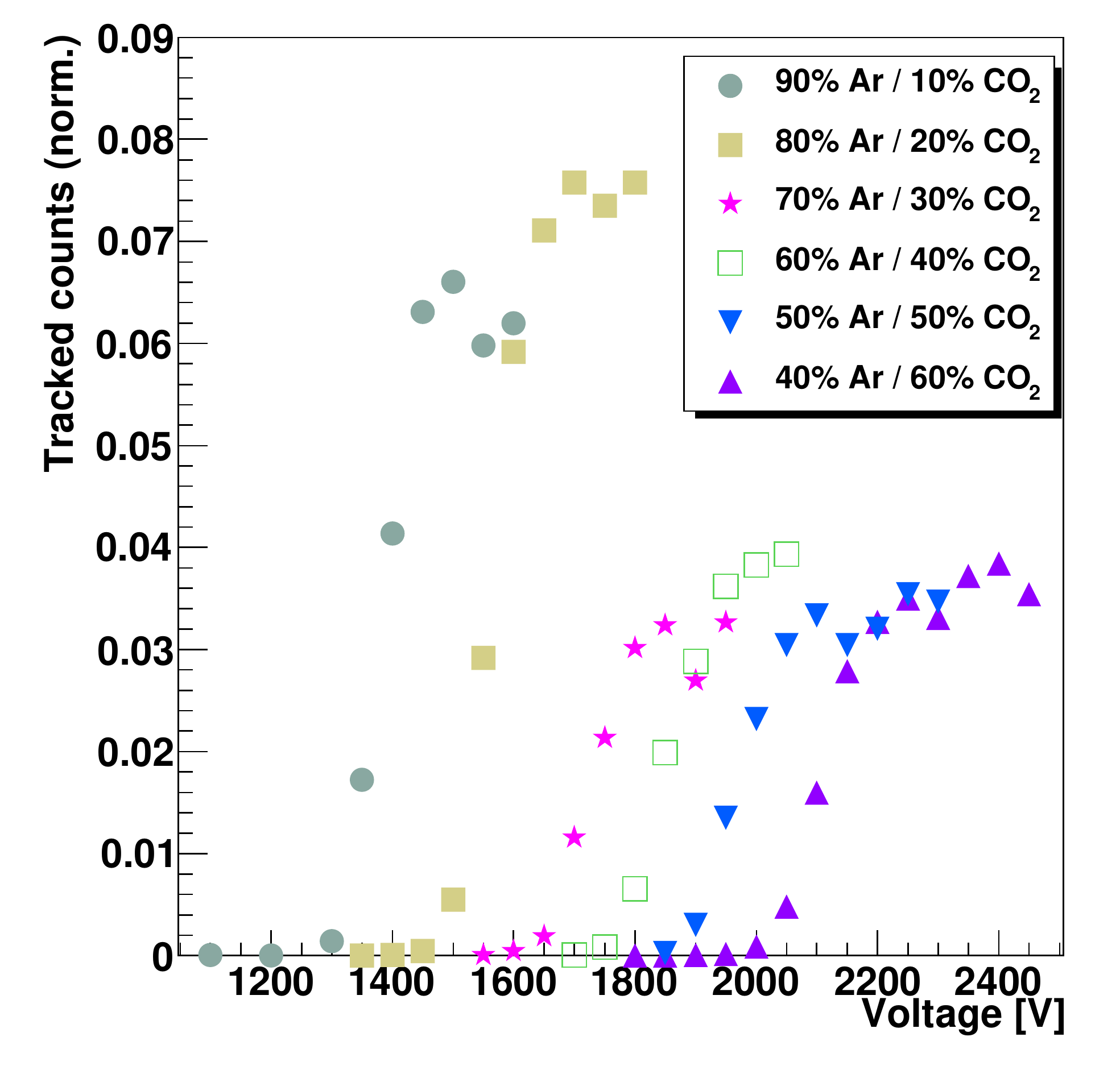}
\end{tabular}
\caption[]{\label{effic} Cosmic ray detection efficiency as a function of high voltage for several argon/CO$_2$ gas mixtures.}
\end{figure}
Cosmic ray data were taken at different high voltage settings and gas mixtures. Tracks were reconstructed to count the number of cosmic rays that passed through a given straw. That number is normalized to the total number of cosmic triggers. The results for different argon-CO$_2$ mixtures as a function of operating voltage are shown in Fig.~\ref{effic}. One can clearly see that the onset of the plateau shifts to lower voltages as the argon concentration in the gas mixture increases. The nominal operating voltage for a given gas mixture is chosen to be slightly above the onset of the plateau to be insensitive to small voltage variations but minimize the dark current at the same time. The different plateau heights for the different gas mixtures are due to the fact that the parameters for drift time (see section~\ref{driftt}) and time-to-distance relations (see section~\ref{postt}), which are used by the tracking algorithm, are not optimized for each gas mixture separately. The working voltages for the different gas mixtures are shown in Table~\ref{workvolt}. The width and position of the pedestal was also studied as a function of bias voltage and no significant changes were observed. 
\begin{table}[h]
\begin{center}
\begin{tabular}{|c|c|}
\hline
{\bf Gas [\% Argon - \% CO$_2$]} & {\bf Working voltage [V]} \\
\hline
90-10 & 1450 \\
80-20 & 1700 \\
70-30 & 1800 \\
60-40 & 1950 \\
50-50 & 2100 \\
40-60 & 2250 \\
\hline
\end{tabular}
\caption{Operating voltages for several gas mixtures.}
\label{workvolt}
\end{center}
\end{table}

\subsection{Gas gain}
The MC simulation described in section~\ref{MCS} can be used to determine the detector gas gain as a function of gas mixture. The detector response to $^{55}$Fe spectra can be simulated and fit to real data. The only free parameter is the gas gain. Table~\ref{gasgains} shows gas gain for different applied voltages and gas mixtures, measured at voltages lying on the plateau. Gas gain between 10$^5$ and 10$^6$ is observed.
\begin{table}[h]
\begin{center}
\begin{tabular}{|c|c|c|}
\hline
{\bf Gas [\% Argon - \% CO$_2$]} & {\bf Voltage [V]} & {\bf Gas gain [10$^5$]}\\
\hline
 90-10  & 1450 & 1.5\\
 80-20 & 1700 & 1.4\\
 70-30 & 1750 & 1.6\\
 70-30 & 1800 & 1.7\\
 60-40 & 1900 & 1.7\\
 50-50 & 2100 & 3.1\\
 40-60 & 2150 & 1.5\\
\hline
\end{tabular}
\caption{Gas gains for different gas mixtures and high voltage settings.}
\label{gasgains}
\end{center}
\end{table}

\subsection{Drift times}
\label{driftt}
An important input parameter for the tracking algorithm is the time difference between the initial ionization and the arrival time of the first electron, referred to as drift time. This is the time which the ionization electron generated closest to the wire needs to drift to the wire. The drift time is obtained from the flash ADC instead of a time-to-digital converter.
 The flash ADC samples the signal at a rate of 125~MHz,
recording 512 time samples of 8~ns, where each time sample contains the measured signal amplitude
digitized in a 12 bit word (the trigger prompting readout and storage of the ADC data is timed to ensure
that the signal data are placed well within this 4096~ns time window). The rising edge of the signal is located as follows:
A signal threshold is calculated from the recorded pedestal values as pedestal mean plus 10 times pedestal
width, and the first time sample where the signal exceeds this threshold is then found and used as a
starting point to determine the rising edge.
The algorithm scans successively earlier time samples until a local ADC minimum is reached, and
successively later samples until a local ADC maximum is reached. These two samples are then taken as the
start and the finish of the leading edge.
The time samples forming part of the leading edge are then examined to enable a reduced number of time
samples to be selected for use with the DCOG method. Selection is based on the the difference in ADC
value between each time sample and those of its neighbors.
For time sample $i$, $d_i$ is the increase in ADC value, $ADC_i$, from that of the previous time sample:
\begin{equation}
  d_i = ADC_i - ADC_{i-1}.
\end{equation} 

The maximum difference ($d_{max}$) is the maximum value of $d_i$ and is located (in time) in the middle of the leading edge. Only time samples that have $d_i$ larger than $f \cdot d_{max}$ are accepted as part of the leading edge and passed to the DCOG method. It was empirically found that $f = 0.3$ produces the best tracking resolutions. If it turns out that this algorithm is too complex to be programmed in the flash ADC, $f$ will be set to zero. This will result in a position resolution that is about 8~$\mu$m worse compared to $f=0.3$.  

\begin{figure}
\begin{tabular}{cc}
  \includegraphics[width=1.0\linewidth]{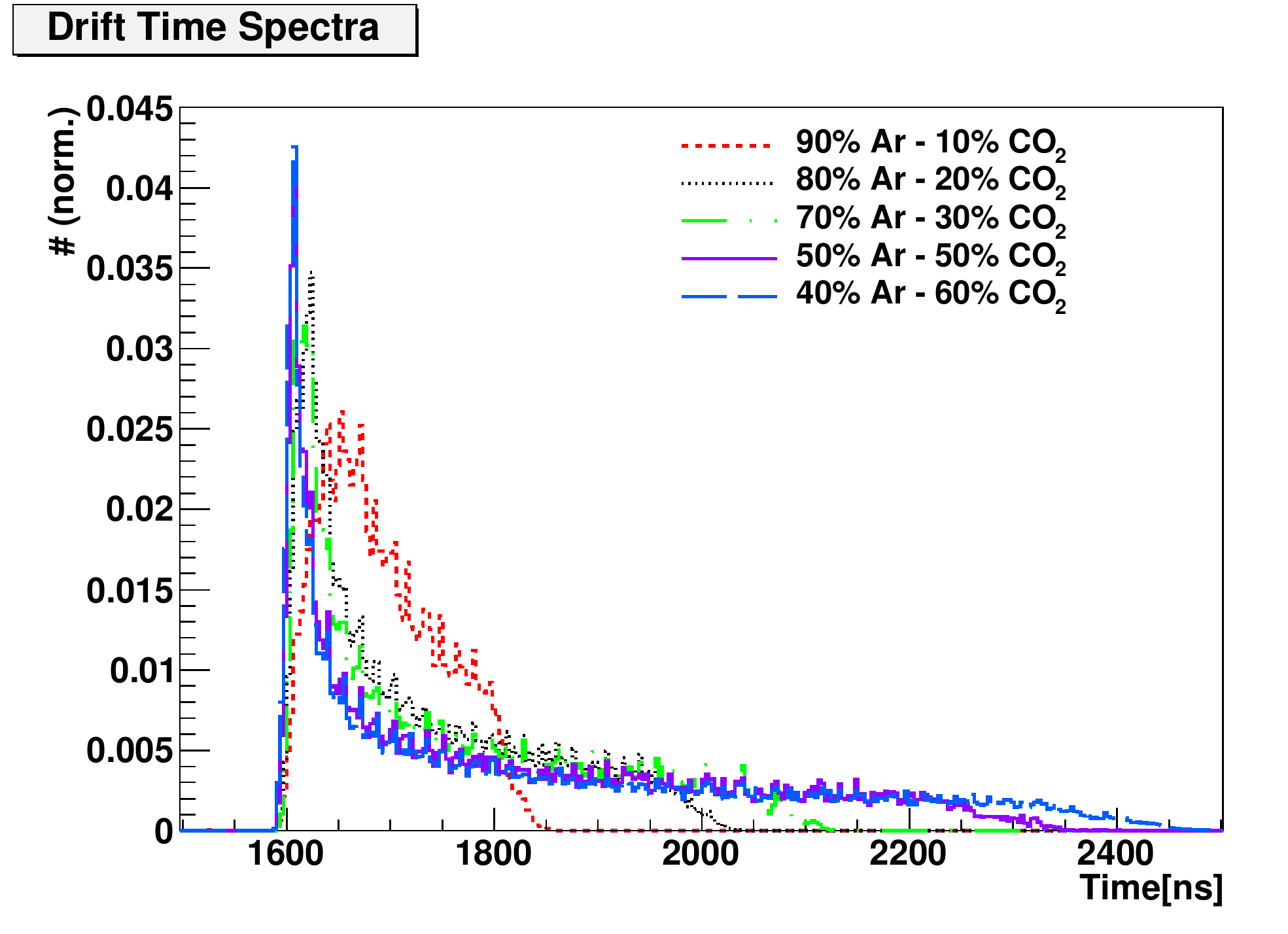}
\end{tabular}
\caption[]{\label{drifttimes} Drift time spectra for different argon-CO$_2$ gas mixtures.}
\end{figure}

Using only the selected samples of the leading edge the extracted drift time ($t_d$) equals:

\begin{equation}
  t_d = \frac{\sum^{n}_{i=2} d_i\cdot t_i}{\sum^{n}_{i=2} d_i},
\end{equation}  
with $n$ the number of samples of the leading edge that are used and $t_i$ is the time stamp of time-sample $i$. There are other methods to extract the drift time, such as the First Electron Method (FEL) where a line if fit to the leading edge, but using the DCOG method resulted in the best resolution. Drift time histograms for different gas mixtures are shown in Fig.~\ref{drifttimes}.
One can observe clear differences between gas mixtures. The more CO$_2$ is added, the longer the drift time becomes. The ``rising edge'' of the drift time histogram is fit with a linear function to extract the lowest possible value for the drift time $t_0$, that is when the ionization-electron is generated very close to the wire. The actual drift time is then $t_d-t_0$. 

The upper limit for the drift time corresponds to the drift distance between the straw wall and the wire. This time should be kept below 1~$\mu$s in order to minimize pile-up. In Table~\ref{drifttimedata} the maximum drift times are listed for the various gas mixtures that were tested. The maximum drift time increases with increasing CO$_2$ content and reaches 900~ns for the 40\%-60\% argon-CO$_2$ gas mixture without the presence of an external magnetic field. See section~\ref{extfield} for a discussion for the response in the magnetic field environment of the experiment.

\begin{table}[h]
\begin{center}
\begin{tabular}{|c|c|c|}
\hline
{\bf Gas} & \multicolumn{2}{c|}{\bf Maximum drift time~[ns]} \\
\cline{2-3}
\bf{ [\% Argon - \% CO$_2$]}& {\bf B = 0~T} & {\bf B = 2.24~T} \\ 
\hline
90-10 & 250 & N/A \\
80-20 & 420 & N/A \\
70-30 & 480 & 620 \\
60-40 & 590 & 670 \\
50-50 & 750 & 790 \\
40-60 & 900 & 920 \\
\hline
\end{tabular}
\caption{List of the maximum drift time for the gas mixtures that were tested without an external magnetic field and simulated with an external magnetic field (see section~\ref{extfield}).}
\label{drifttimedata}
\end{center}
\end{table}

\subsection{Position resolution}
\label{postt}
\begin{figure}\begin{tabular}{cc}
\includegraphics[width=0.75\linewidth]{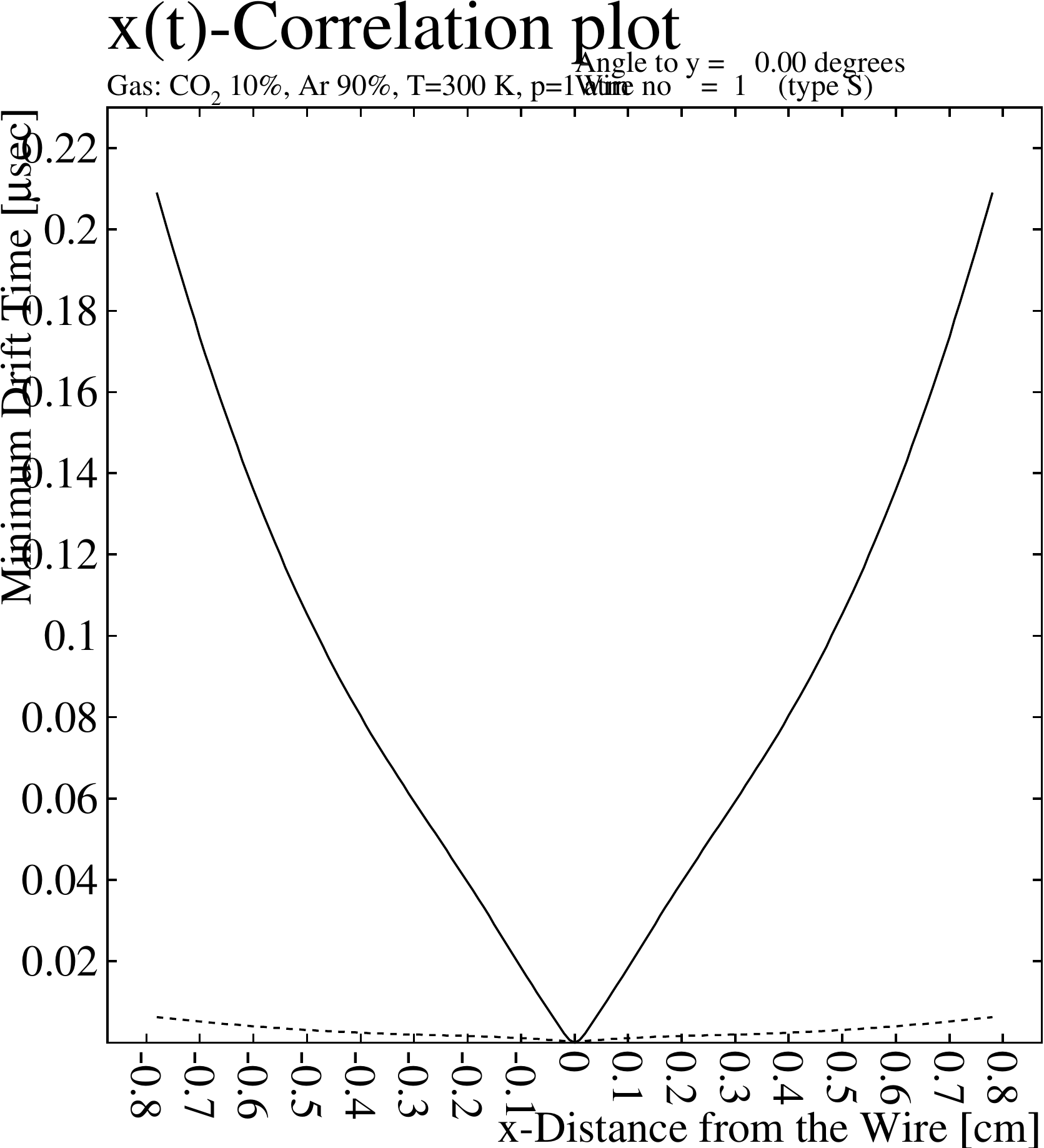}\\
\includegraphics[width=0.75\linewidth]{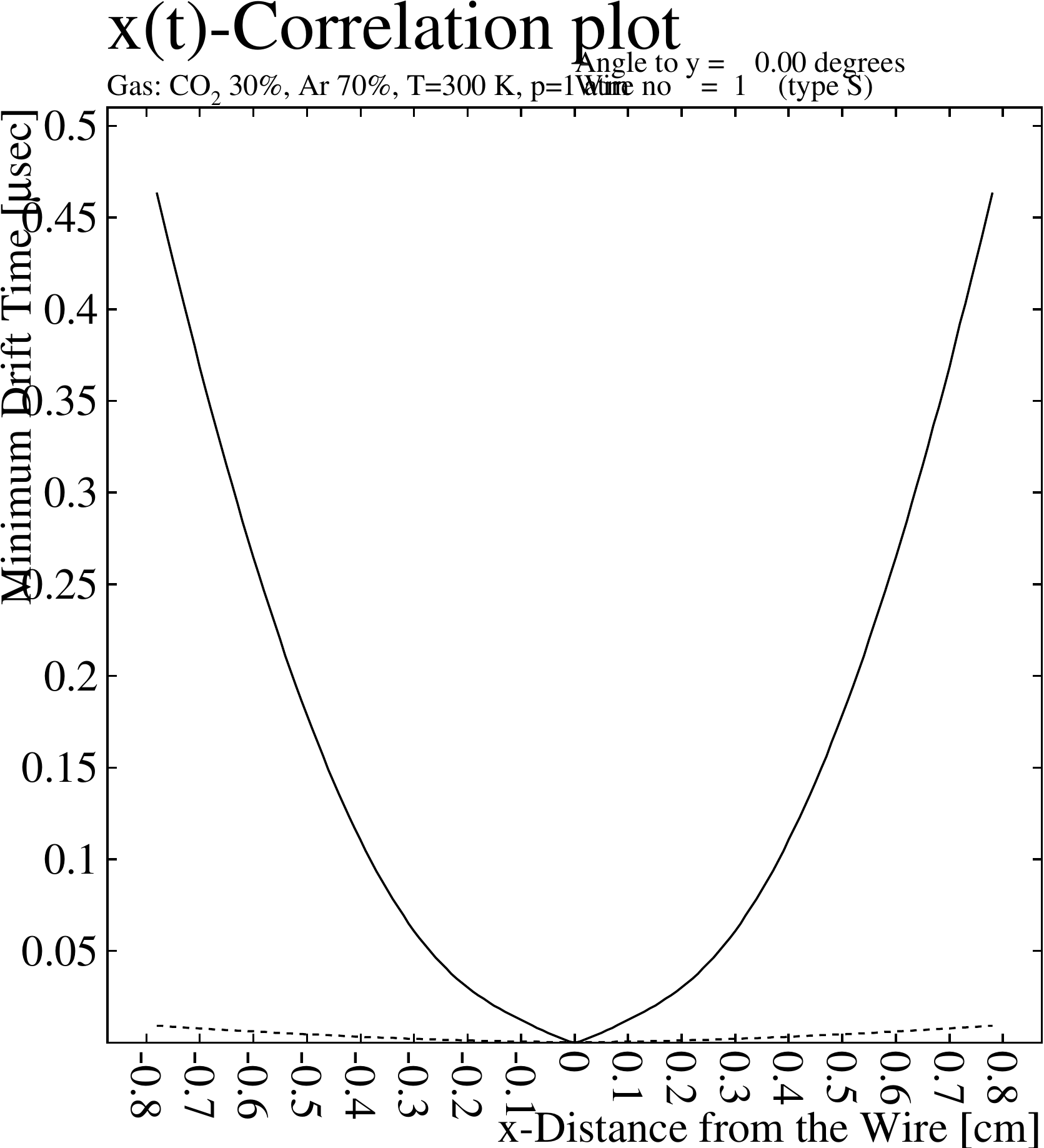}
\end{tabular}
\caption[]{\label{t2r} Drift-to-distance relationship for two gas mixtures with no external magnetic field applied. Left: 90\%-10\% argon-CO$_2$ (1450~V on the wires); right: 70\%-30\% argon-CO$_2$ (1750~V on the wires).}
\end{figure}

Once the actual drift time ($t_d-t_0$) is extracted, it is converted into a distance from the wire (radius). This conversion is performed using time-to-distance tables that are calculated using {\sc Garfield}. In Fig.~\ref{t2r} this relationship is shown for two gas mixtures with no external magnetic field applied. Once drift radii are extracted, they can be used in the track finding/fitting algorithm and resolutions can be calculated. Tracks are defined in the plane perpendicular to the wires. An illustration of how the tracking works is shown in Fig.~\ref{smallprot2}. The brown drift-circles are circles defined by the extracted drift radii. If there are more than three hits, a track can be fit: this is shown as the blue line.  

\subsubsection{Track finding}
When a cosmic ray passes through the detector, it can traverse through multiple straws. An electrical signal is induced on the wire in each straw which the cosmic ray passed through. From timing characteristics of this signal the distance of closest approach can be reconstructed (hit-radius). When 3 or more hit-radii are reconstructed in an event, a track finding algorithm is applied. Every combination of two hit-radii is taken
to provide four possible tracks by tangent lines connecting the two circles defined by the two hit-radii. These trajectories are then used to search other straws with hit-radii within a 4~mm window of these track candidates. The track with the most hits or the lowest $\chi^2/dof$ (if the number of hits is the same) is identified as the actual track and passed to the track fitting algorithm.

\subsubsection{Track fitting}
At this stage, the distances between the found track in the $x$-$y$ plane (plane perpendicular to the axial wires) with equation: $Ax+By+1=0$ and the hit-radii are minimized using {\sc Minuit}~\cite{MINUIT}. This software package finds the values for $A$ and $B$ which minimizes the function:

\begin{equation}
  \sum^{\#hits}_{i=1}\left[ \frac{r_{hit}(i)^2- \frac{\left( Ax_0\left(i\right)+By_0\left(i\right)+1\right)^2}{A^2+B^2}}{\sigma_{hit}(i)} \right]^2.
\end{equation} 
The numerator is the difference between the hit radius squared ($r_{hit}^2$) and the distance between the found track and the center of the straw squared ($\frac{\left( Ax_0+By_0+1\right)^2}{A^2+B^2}$), with ($x_0$,$y_0$) the coordinates of the center of the straw. The error on the hit radius, $\sigma_{hit}$, is set to a constant and does not have an influence on the minimization.

\subsubsection{No external magnetic field}
\begin{figure}
\begin{tabular}{cc}
  \includegraphics[width=0.49\linewidth]{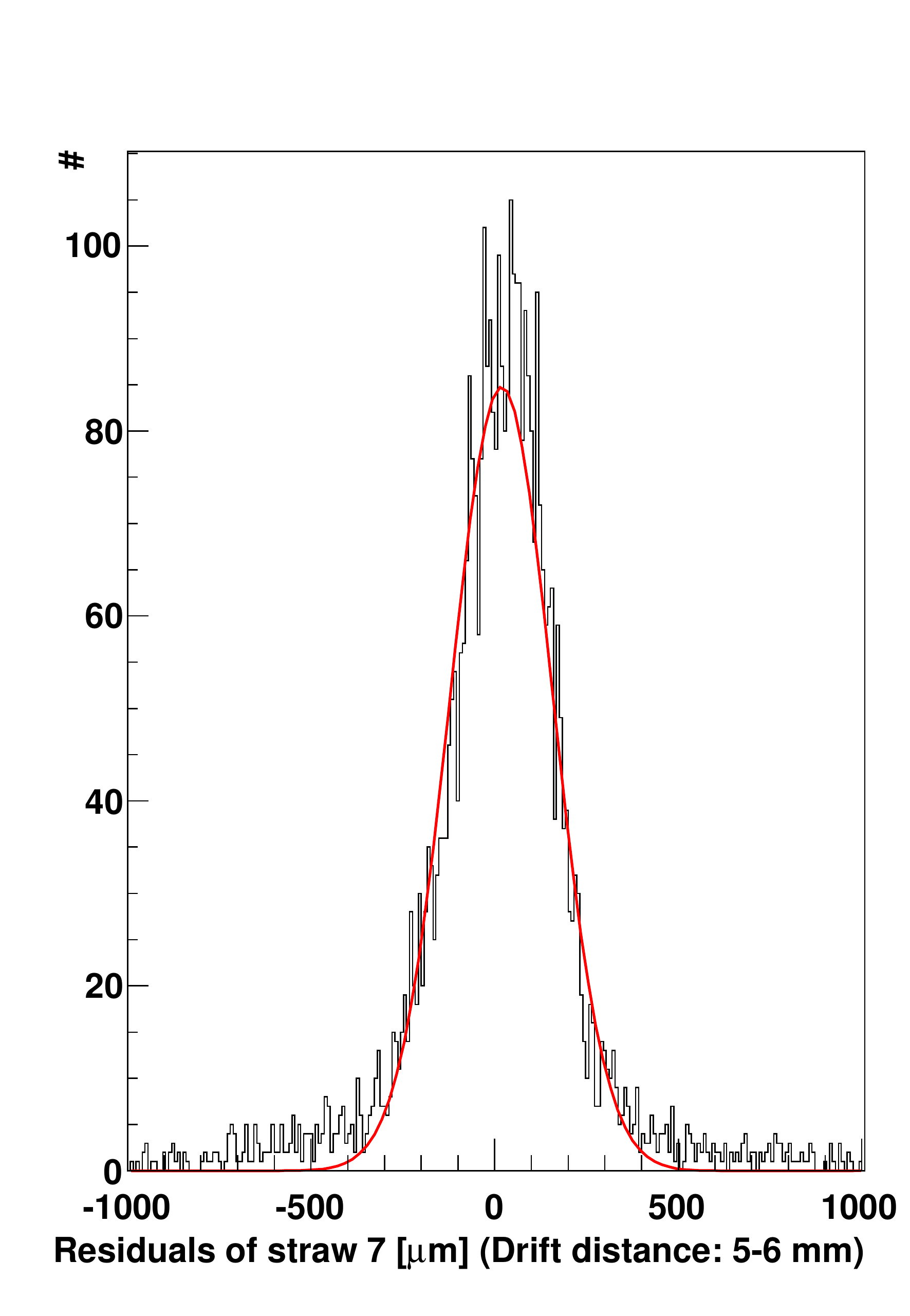}
  \includegraphics[width=0.49\linewidth]{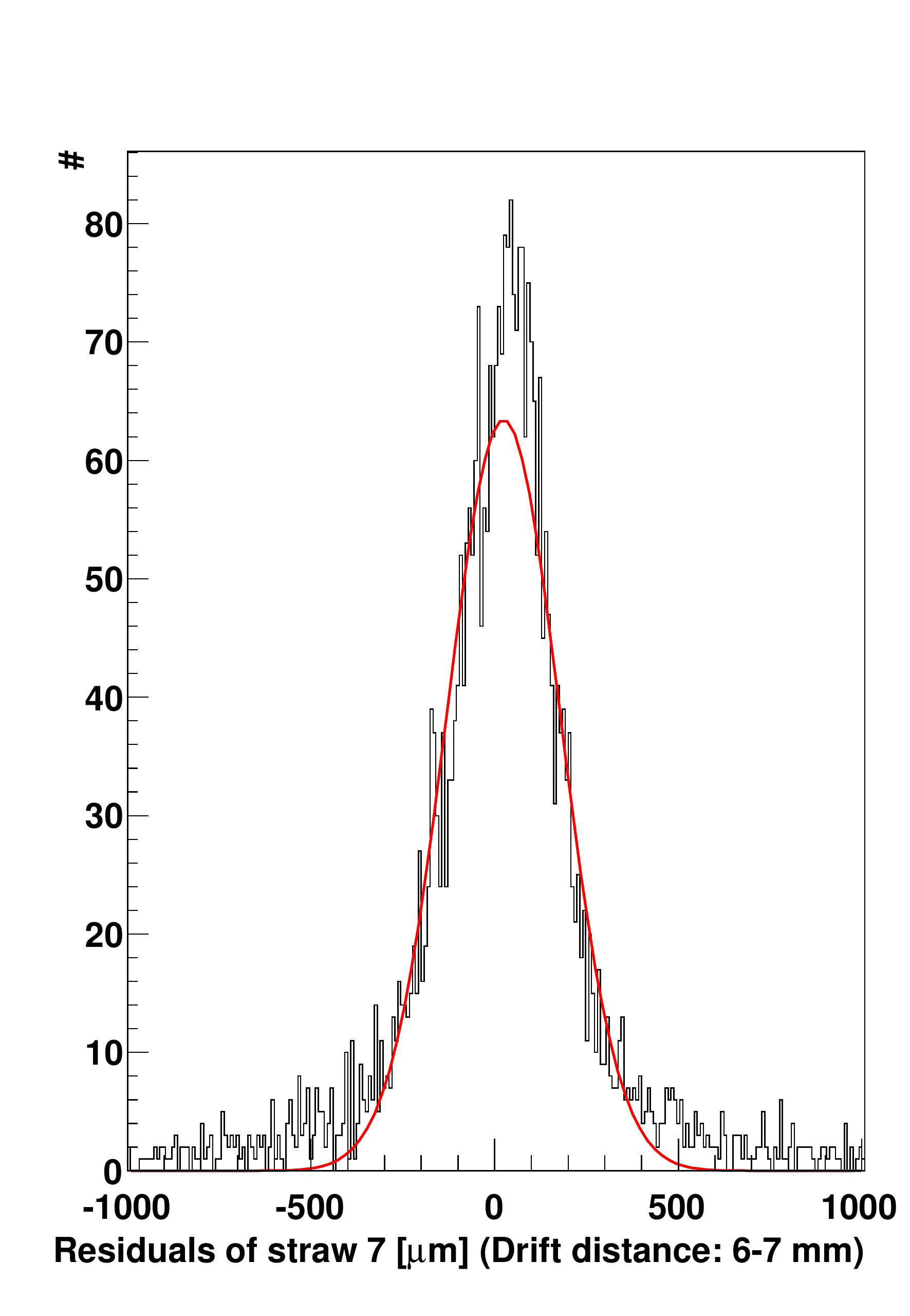}
\end{tabular}
\caption[]{\label{residex} Two examples of residual histograms. The red line indicates the result of a Gaussian fit.}
\end{figure}

The position resolution is calculated for wire 7 for different gas mixtures as a function of the straw radius. This wire was chosen because of its central location. This calculation is done by selecting tracks with at least 5 hits and performing track fitting on them with wire 7 left out. The distance between the hit-radii of wire 7 and the fit tracks (also called residuals) are put into histograms. Examples of residual histograms can be seen in Fig.~\ref{residex}. This plot is fit with a Gaussian and the width of the Gaussian is quoted as the resolution. 

\begin{figure}
\begin{tabular}{cc}
  \includegraphics[width=0.49\linewidth]{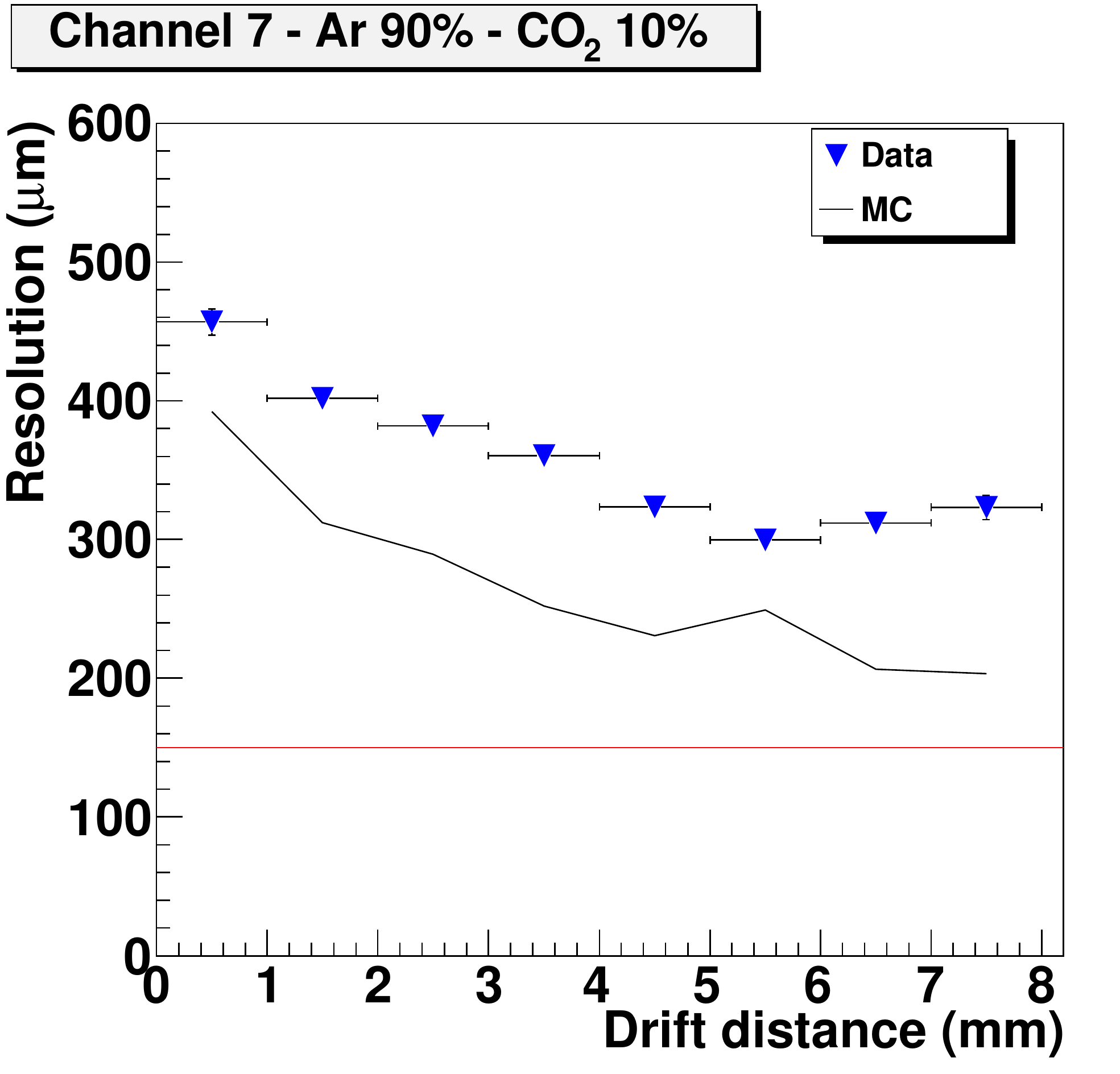}
  \includegraphics[width=0.49\linewidth]{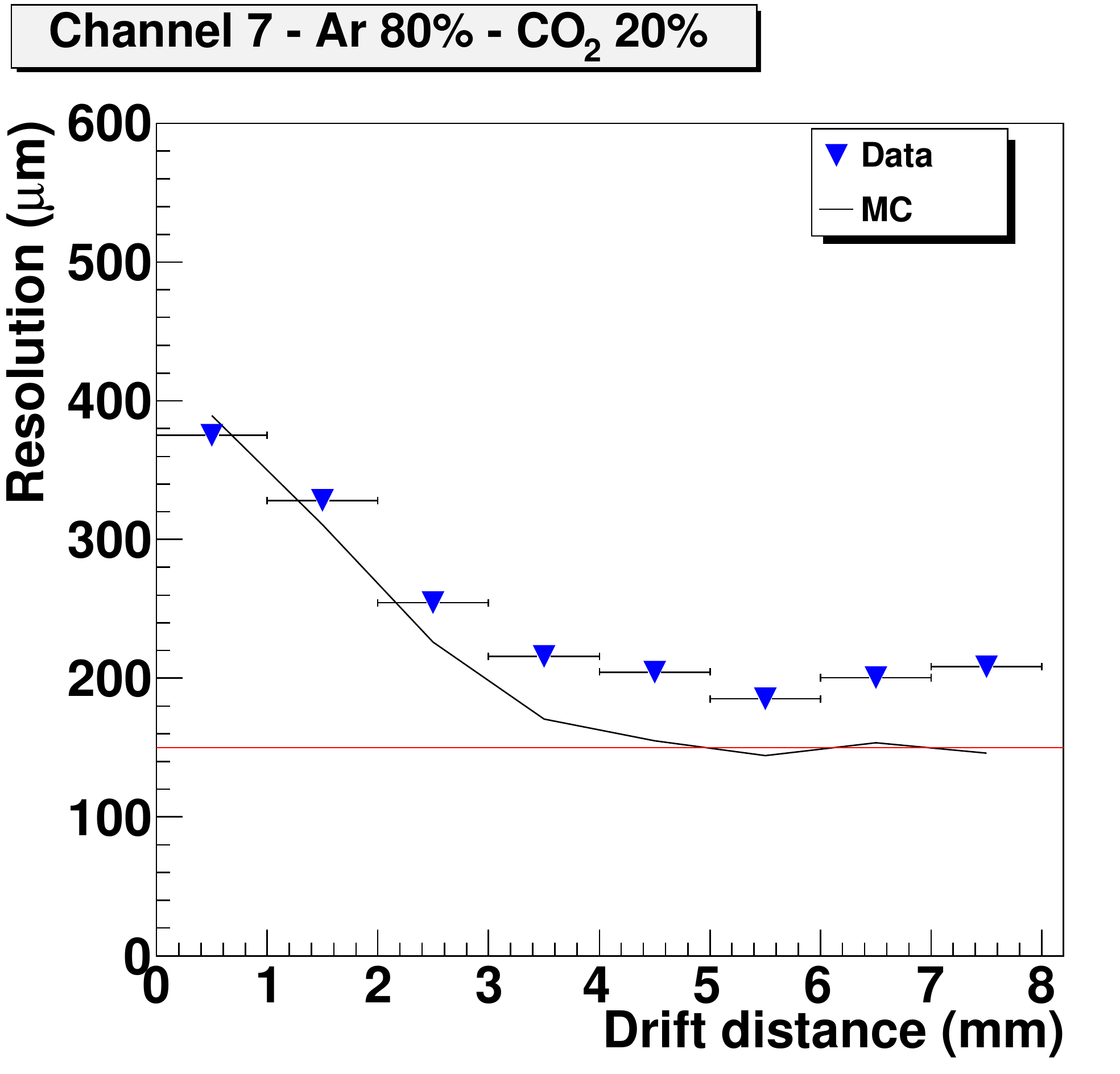}\\
\includegraphics[width=0.49\linewidth]{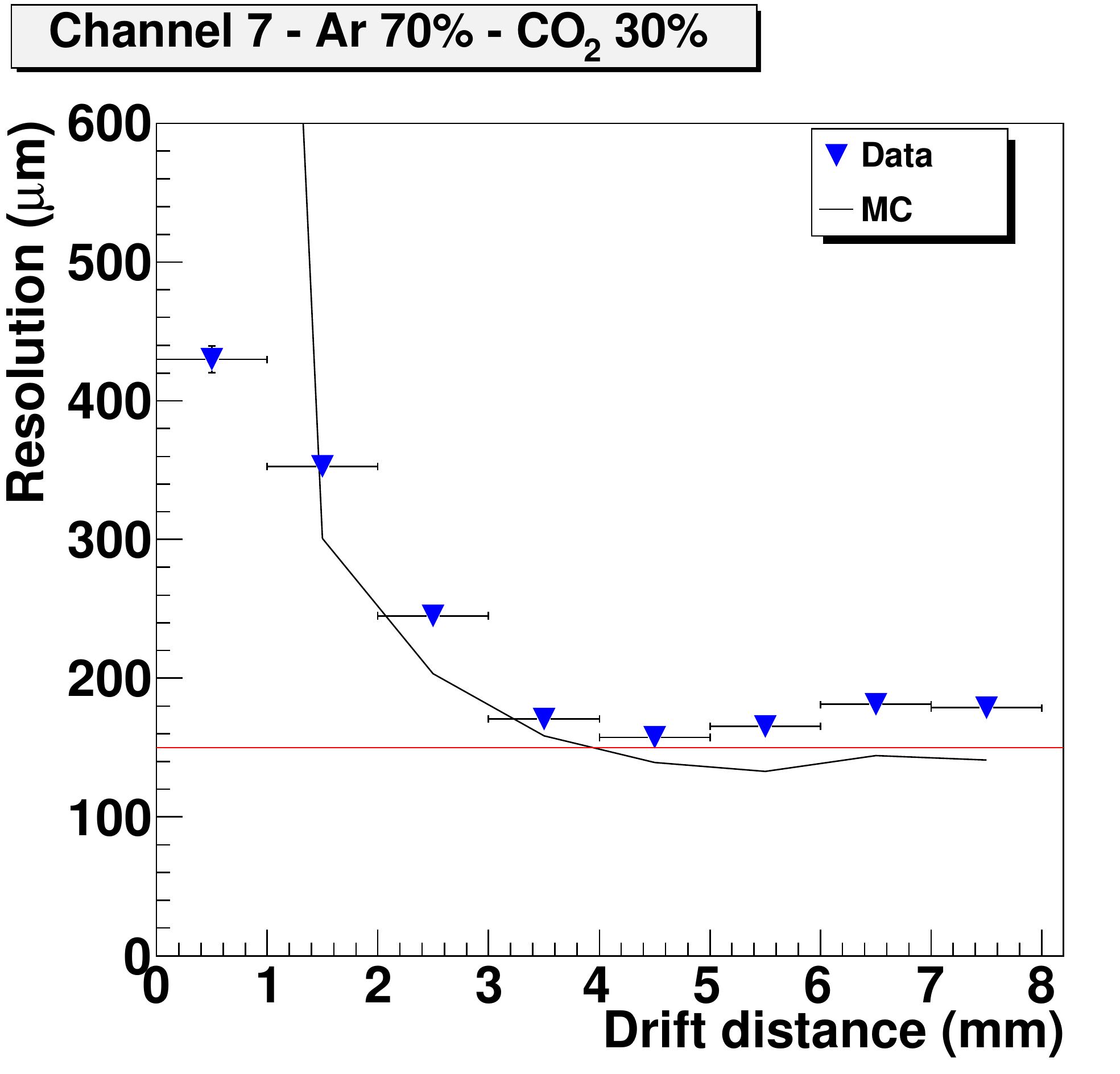}
\includegraphics[width=0.49\linewidth]{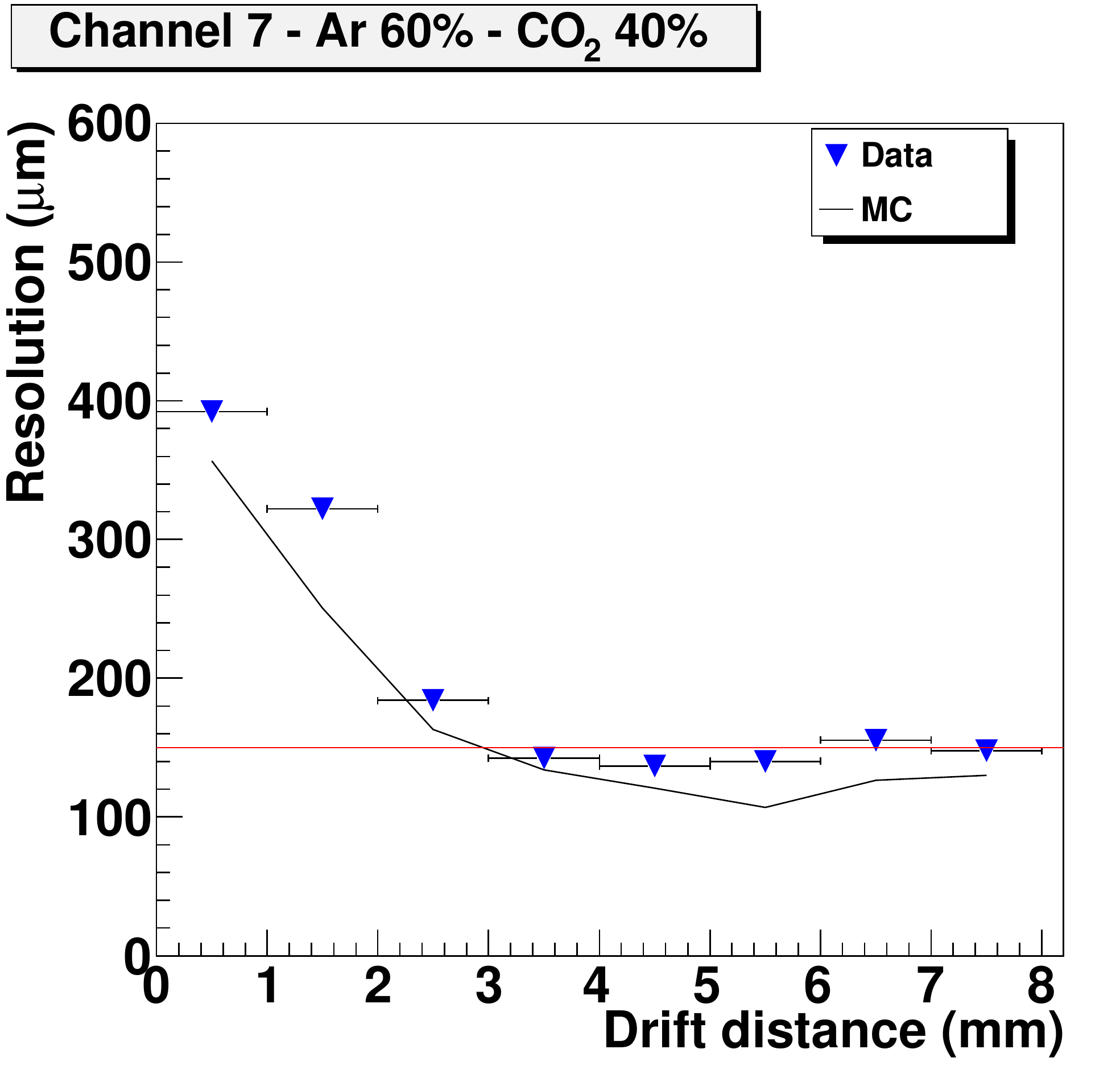}\\
\includegraphics[width=0.49\linewidth]{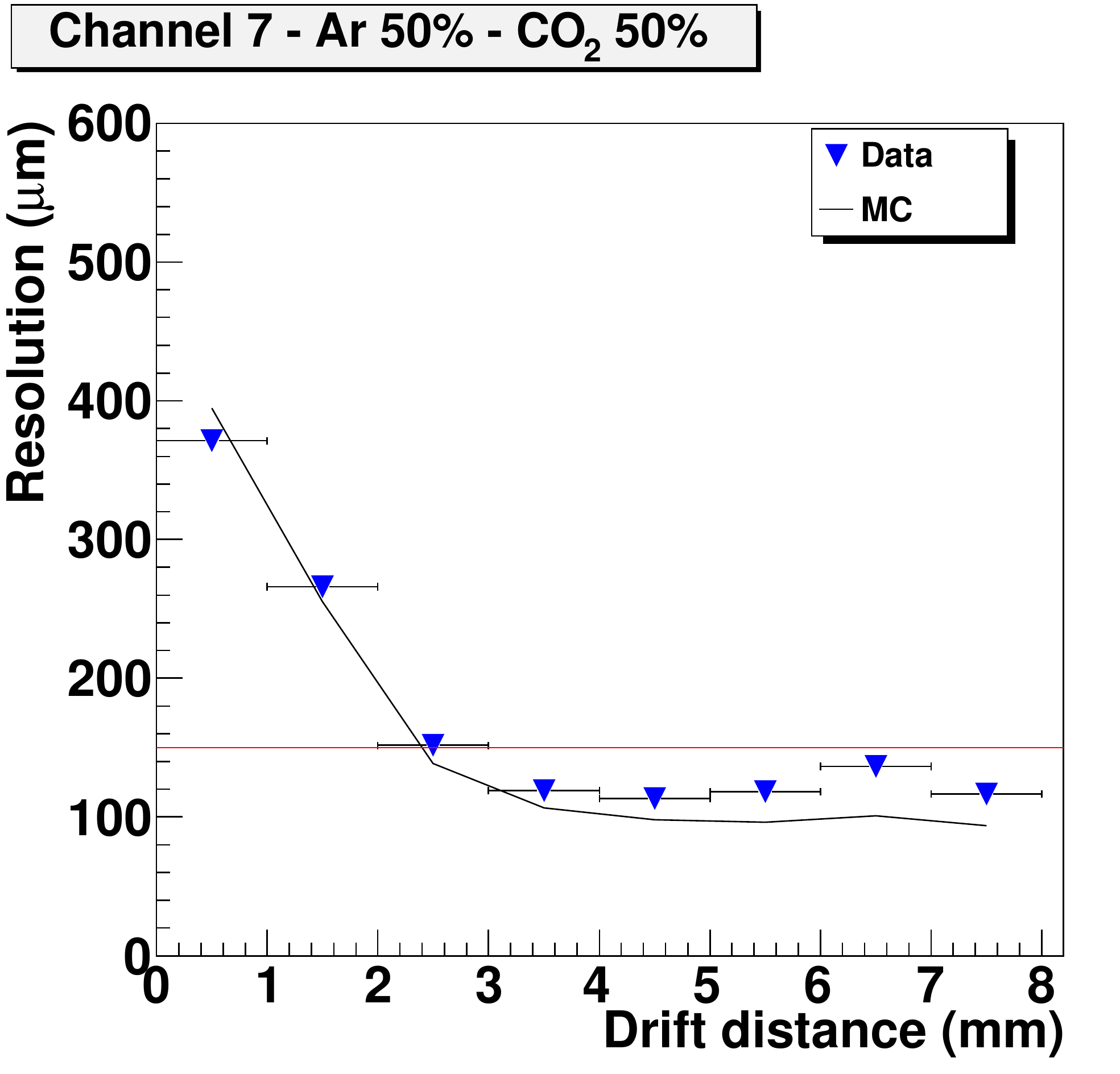}
  \includegraphics[width=0.49\linewidth]{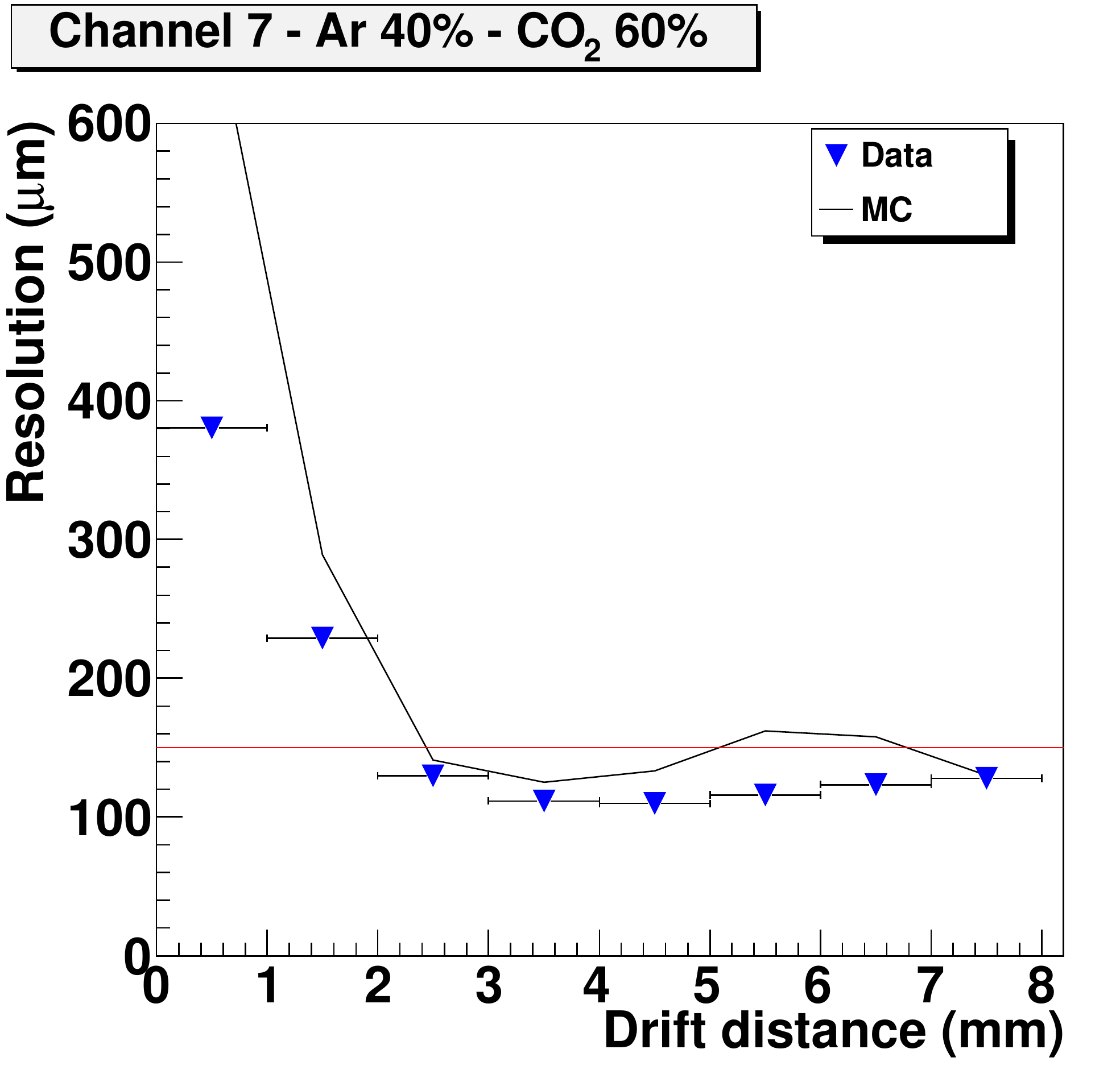}
\end{tabular}
\caption[]{\label{resolutionsdata} Position resolutions as a function of drift distances obtained
using different argon-CO$_2$ mixtures with no external magnetic field present. Both Monte Carlo and data results are shown. The red line indicates the design resolution.}
\end{figure}

The data are compared to MC simulations and shown in Fig.~\ref{resolutionsdata}. It is readily recognized that the resolution depends on the amount of CO$_2$ in the gas mixture. This is due to overall drift times increasing with the amount of CO$_2$ increases. For short drift distances the resolution tends to be worse because the slope in the time-to-distance relation is shallower at drift distances smaller than  3~mm (see Fig.~\ref{t2r}). Generally the MC simulation describes the data very well. This supports the use of the MC to study the effect of an external magnetic field on the position resolution of the detector.

 At very small drift distances the resolution degrades due to discrete ionization statistics close to the wire: the ionization can happen right next to the wire or a little bit above or below it. When the closest ionization does not happen next to the wire the extracted drift radii are too big and this effect causes degrading resolutions.

\subsubsection{External magnetic field}
\label{extfield}
A parameter that changes due to an external magnetic field is the maximum drift time. The longest possible drift times with and without an external magnetic field are shown in Table~\ref{drifttimedata}. Drift times shorter than 1~$\mu$s are desired to avoid pile up. All gas mixtures satisfy this requirement, although the 40\%-60\% argon-CO$_2$ mixture comes close to that limit.

\begin{figure}
\begin{tabular}{cc}
  \includegraphics[width=1.0\linewidth]{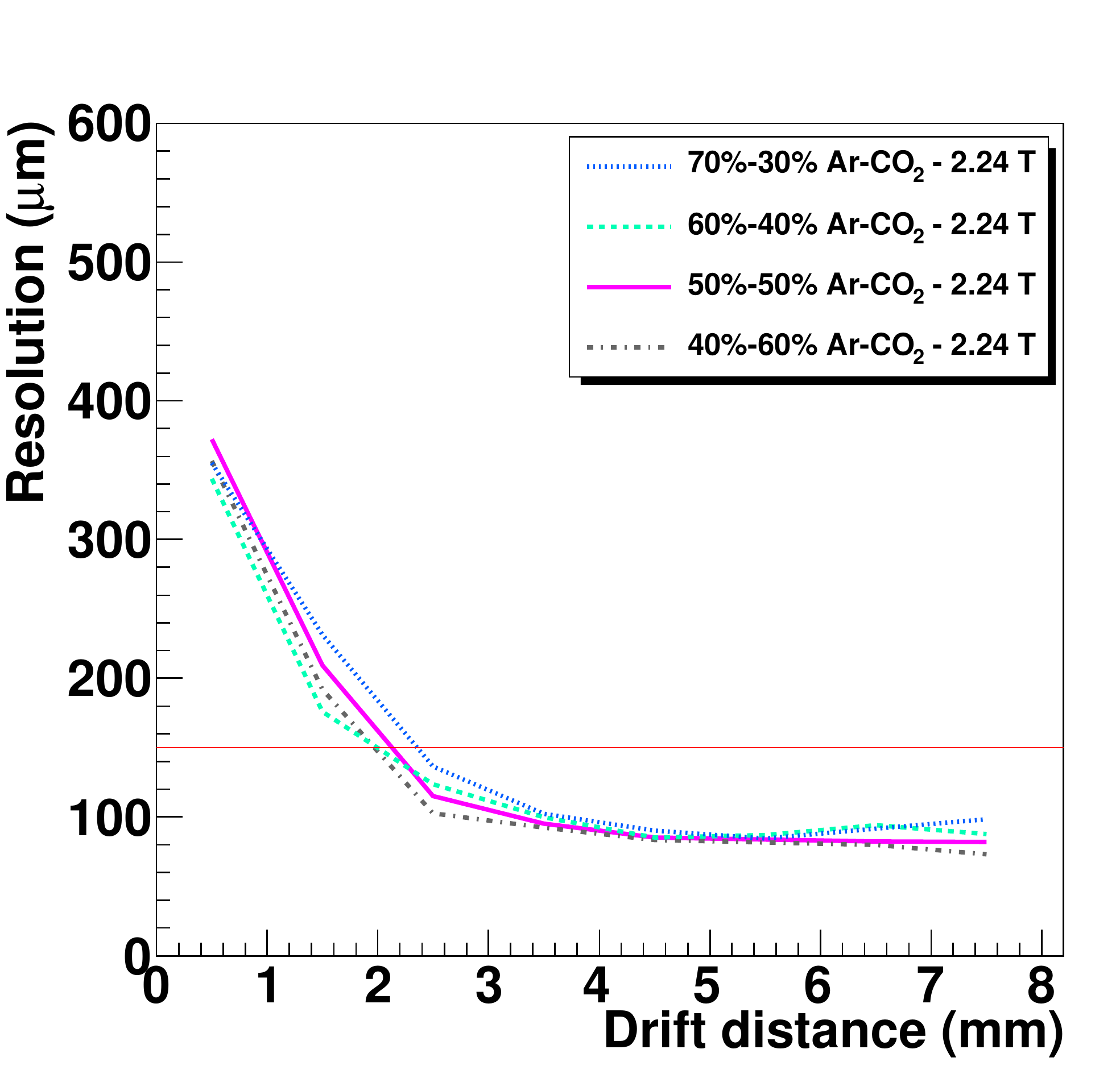}
\end{tabular}
\caption[]{\label{resolutionsmcmag} Position resolution as a function of drift distance obtained
using a Monte Carlo simulation with different argon-CO$_2$ mixtures and a 2.24~T external magnetic field. The red line indicates the design resolution.}
\end{figure}

To examine the position resolution at a magnetic field of 2.24~T provided by the GlueX detector, a MC simulation is used. The results for different gas mixtures are shown in Fig.~\ref{resolutionsmcmag}. One can observe the same general behavior as without an external magnetic field but the resolution is improved overall (around 150~$\mu$m) and meets the specifications defined in~\cite{TD} for drift distances longer than 2~mm.

These tests show that gas mixtures between 60\%-40\% and 40\%-60\% argon-CO$_2$ satisfy the requirements outlined in the introduction and can 
therefore be used in the final setup in Hall-D. It is advantageous for safety considerations that the required resolution can be achieved without including a flammable gas component.

\subsection{Efficiency}
\begin{figure}
\begin{tabular}{cc}
  \includegraphics[width=1.0\linewidth]{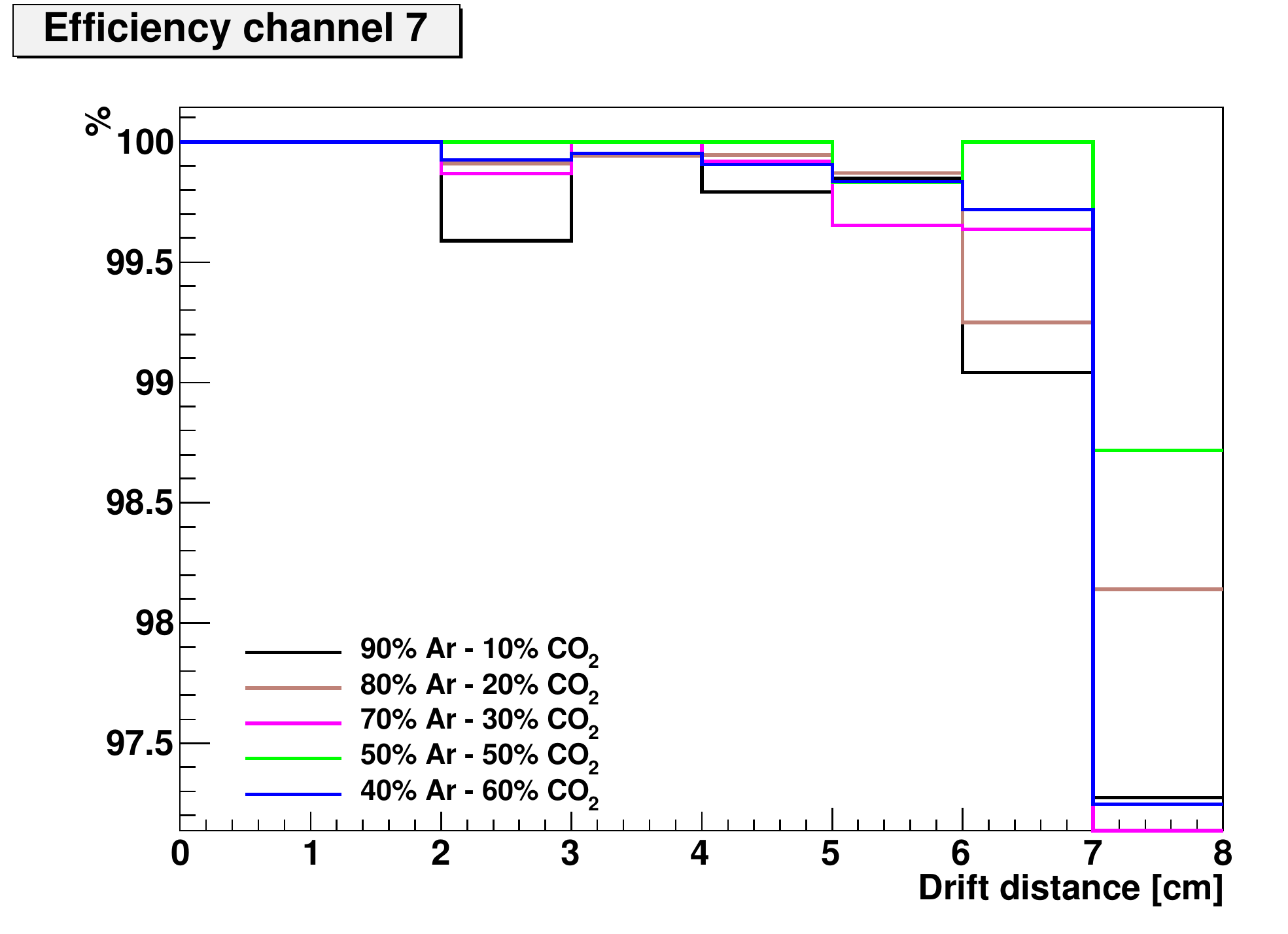}
\end{tabular}
\caption[]{\label{efficiency7} Efficiency of straw 7 as a function of drift distance for several gas mixtures.}
\end{figure}

The efficiency of a given straw is calculated by dividing the number of tracks with five (or more) hits that go through the given straw by the number of tracks with four (or more) hits (given straw excluded) that are expected to go through the given straw. The efficiency of straw 7 as a function of drift distance is shown in Fig.~\ref{efficiency7}. It is close to 100\% but degrades close to the edge of the straw where the track length inside the straw is very short, this can be seen in in Fig.~\ref{smallprot2}: long drift-times mean large drift-circles and short track lengths inside the straw. However, the close-pack design of the CDC will ensure high efficiency for hits in adjacent layers.       
\section{Energy loss studies}
Firstly, the energy loss is studied as a function of incident angle. In order to do this the detector needed to be tilted. The tilt angle is reconstructed from the energy loss. The primary motivation of this study is to see if the tilted detector behaves as expected, because in earlier prototypes deviations were observed that lead back to gas leaks, sagging of the straw, etc. Secondly, the energy loss as a means of identifying low momentum particles is investigated. 
  
\subsection{Tilted chamber}
\begin{figure}
\begin{tabular}{cc}
  \includegraphics[width=1.0\linewidth]{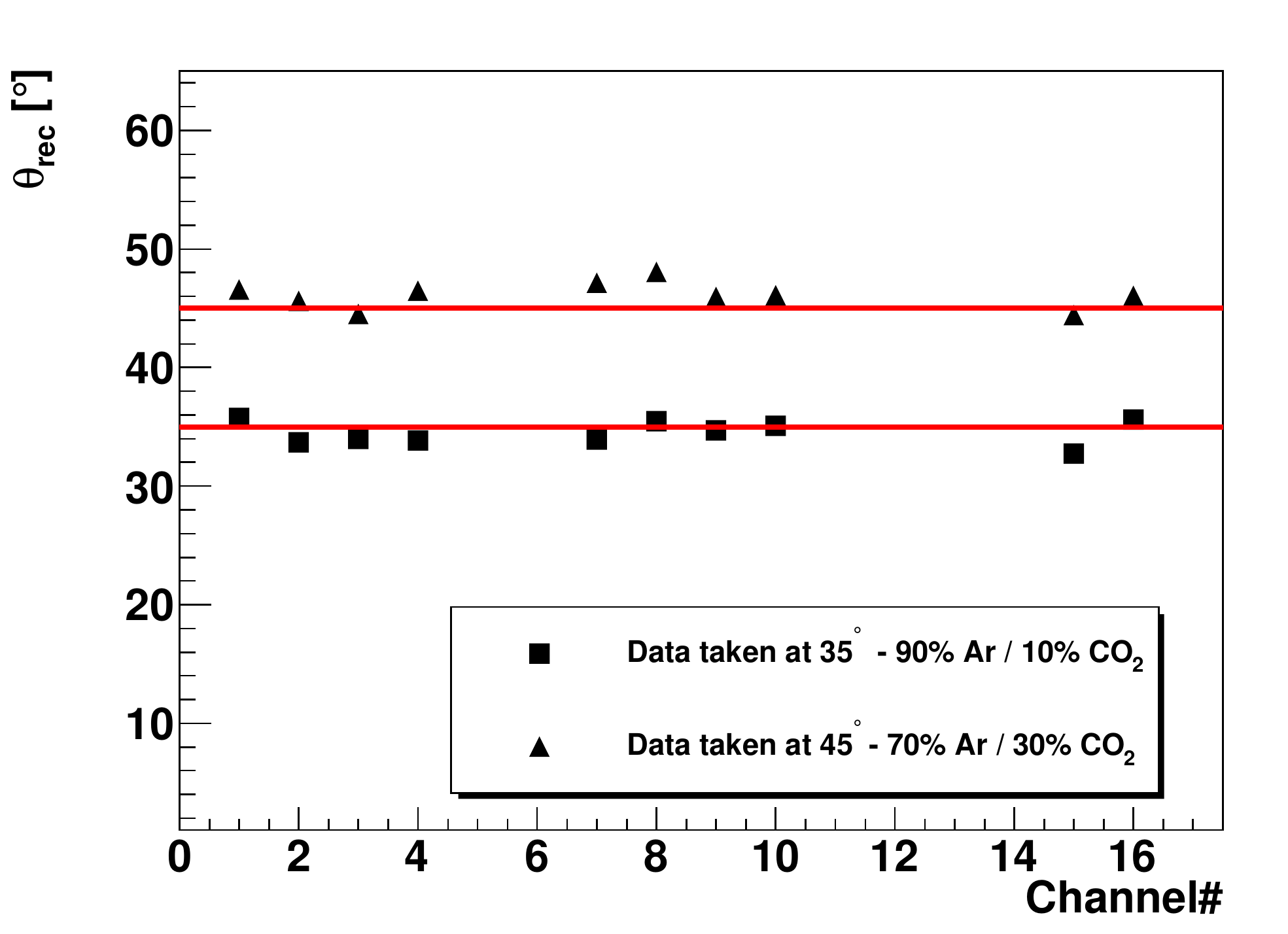}
\end{tabular}
\caption[]{\label{recangle} Reconstructed angle of the tilted chamber as a function of straw number. See text for more details.}
\end{figure}

In order to increase the average incident angle of the cosmic rays with respect to the test chamber, the chamber was tilted as indicated in the top of Fig.~\ref{smallprot1}. In such a configuration the average path length of cosmic rays inside the straw becomes longer and the average energy deposition larger. The chamber was positioned at two different angles: 35 and 45 degrees (see Fig.~\ref{smallprot1}). A 90\%-10\% argon-CO$_2$ gas mixture was used in the 35 degree setup and a 70\%-30\% argon-CO$_2$ mixture in the 45 degree setup. The average energy deposition was measured and compared with the results obtained with the chamber in the horizontal position (see Fig.~\ref{smallprot1}). As a result the measured energy deposition can be used to reconstruct the angle $\theta_{rec}$ to which the chamber was tilted using following relation:

\begin{equation}
  \theta_{rec} = \cos^{-1}\left(\frac{\langle E_0 \rangle \cdot 0.92}{\langle E_{\theta}\rangle}\right).
\end{equation}
$E_0$ is the energy deposition with the detector in horizontal position and $E_{\theta}$ the energy deposition with the detector tilted at the angle $\theta$. Only the energy deposition of fully reconstructed tracks was used. The factor 0.92 is the solid angle correction to the trigger counters, which systematically moves the average track angle with the chamber in its horizontal position. In Fig.~\ref{recangle}, one can see the reconstructed angle as a function of channel number (see Fig.~\ref{smallprot2}). The angles can be accurately reconstructed.  

\subsection{Particle identification}
In order to determine the effectiveness of this chamber to identify pions, kaons, and protons using energy loss (${\rm d}E/{\rm d}x$), a MC study based on the realistic simulation described previously was undertaken. There is no intention to give a complete description of Particle IDentification (PID) here, but rather to investigate the principle. Pions, kaons, and protons were generated using the MC described in section~\ref{MCS}. The ${\rm d}E/{\rm d}x$ or energy ($E$) loss per unit of length ($x$) distribution in one straw is calculated. An example of a ${\rm d}E/{\rm d}x$ distribution in one straw (one layer) is shown in Fig.~\ref{350onehit}). A toy MC is used to generate energy losses according to the extracted distributions. In this way energy losses of multiple hits (or layers) can be combined to improve PID possibilities. When multiple layers are combined, the hit with the largest ${\rm d}E/{\rm d}x$ and the hit with the lowest ${\rm d}E/{\rm d}x$ are removed. This is an empirical method to make the Landau tails shorter and improve the separation of particles. However, in this case this method only slightly improves PID possibilities. 

 \subsubsection{350~MeV Pions, kaons, and protons}
\begin{figure}
\begin{tabular}{cc}
  \includegraphics[width=0.95\linewidth]{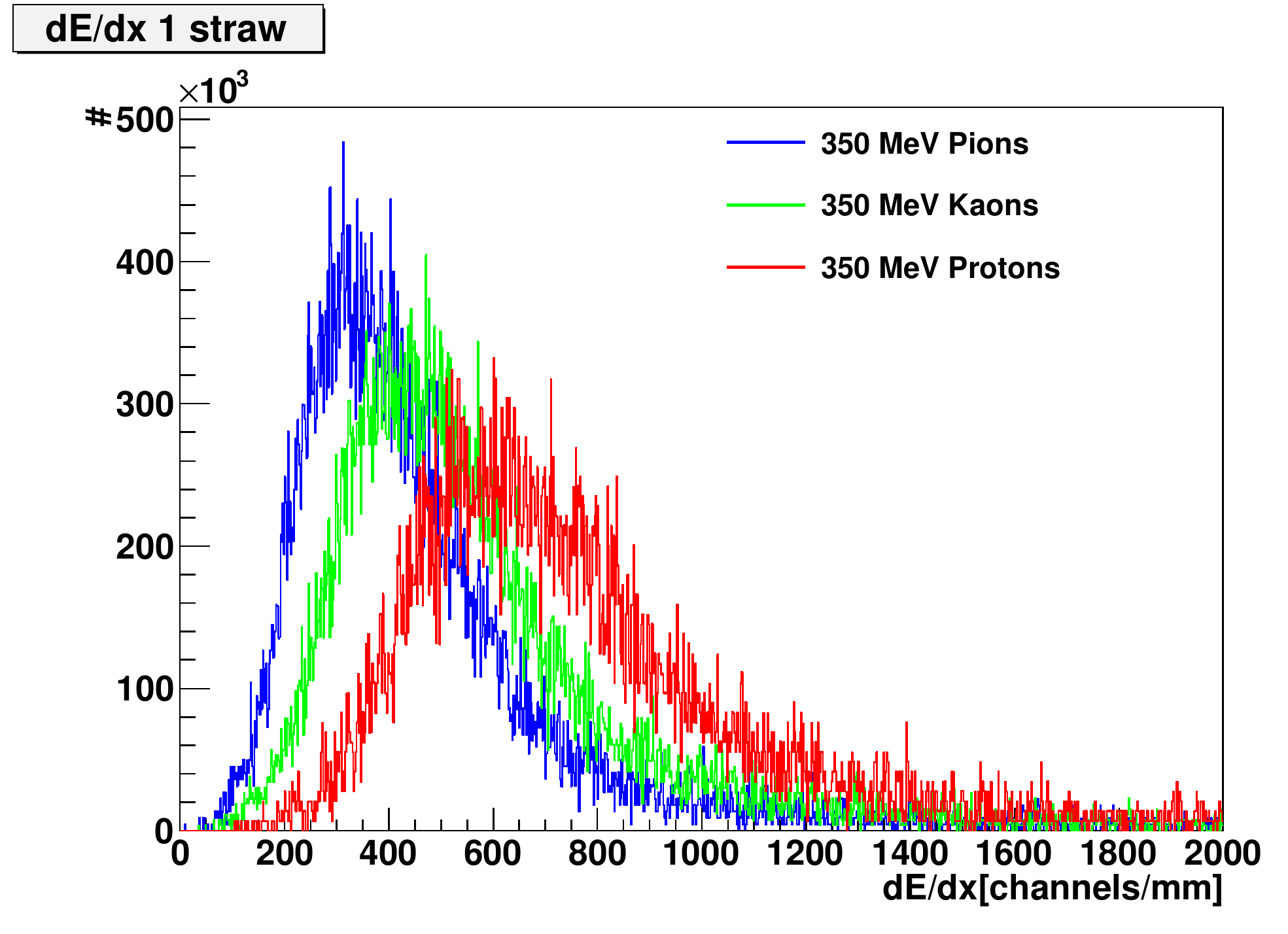}
\end{tabular}
\caption[]{\label{350onehit} Pion, kaon, and proton ${\rm d}E/{\rm d}x$ distribution.}
\end{figure}

\begin{figure}
\begin{tabular}{cc}
  \includegraphics[width=0.5\linewidth]{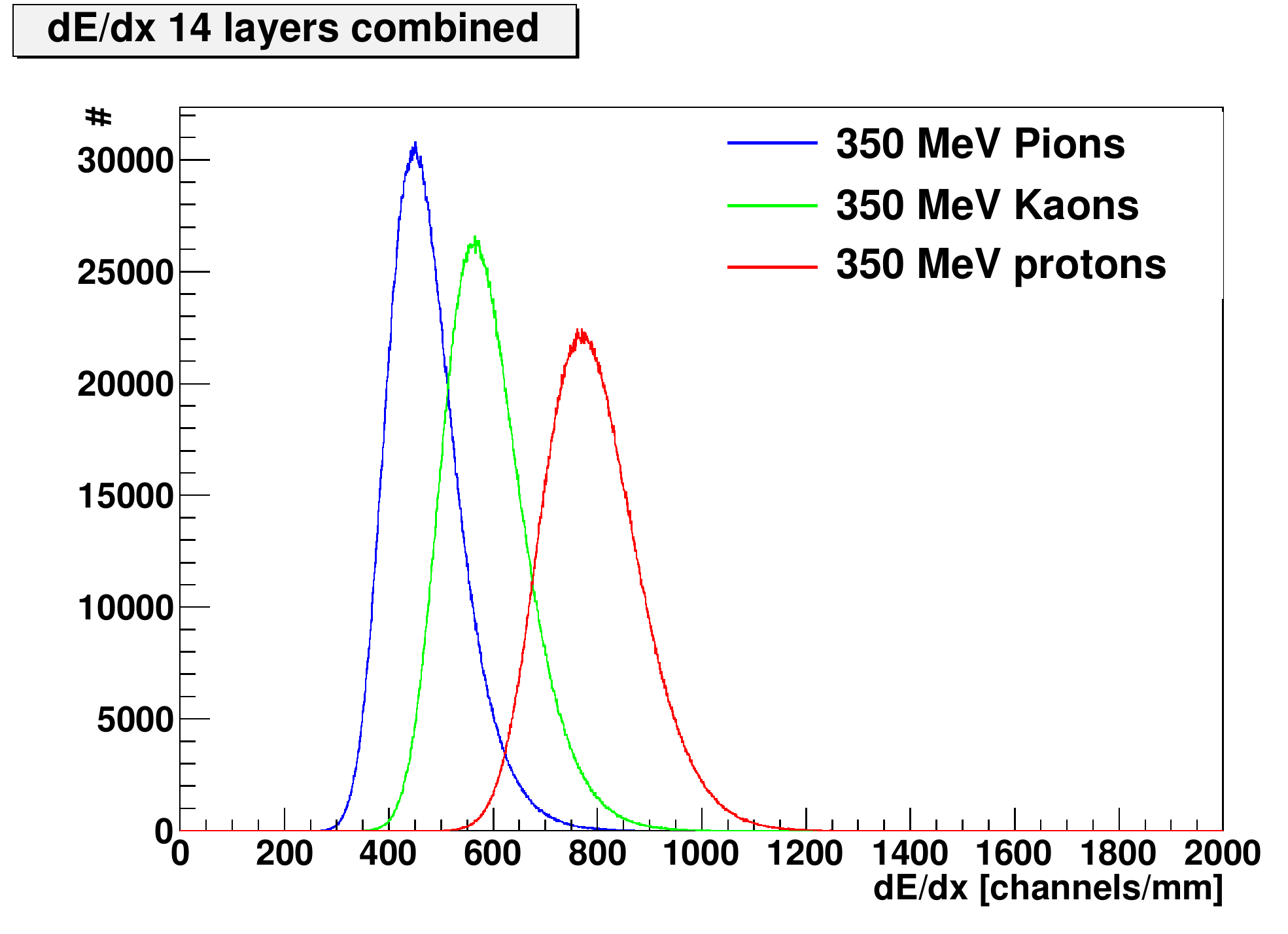}
  \includegraphics[width=0.5\linewidth]{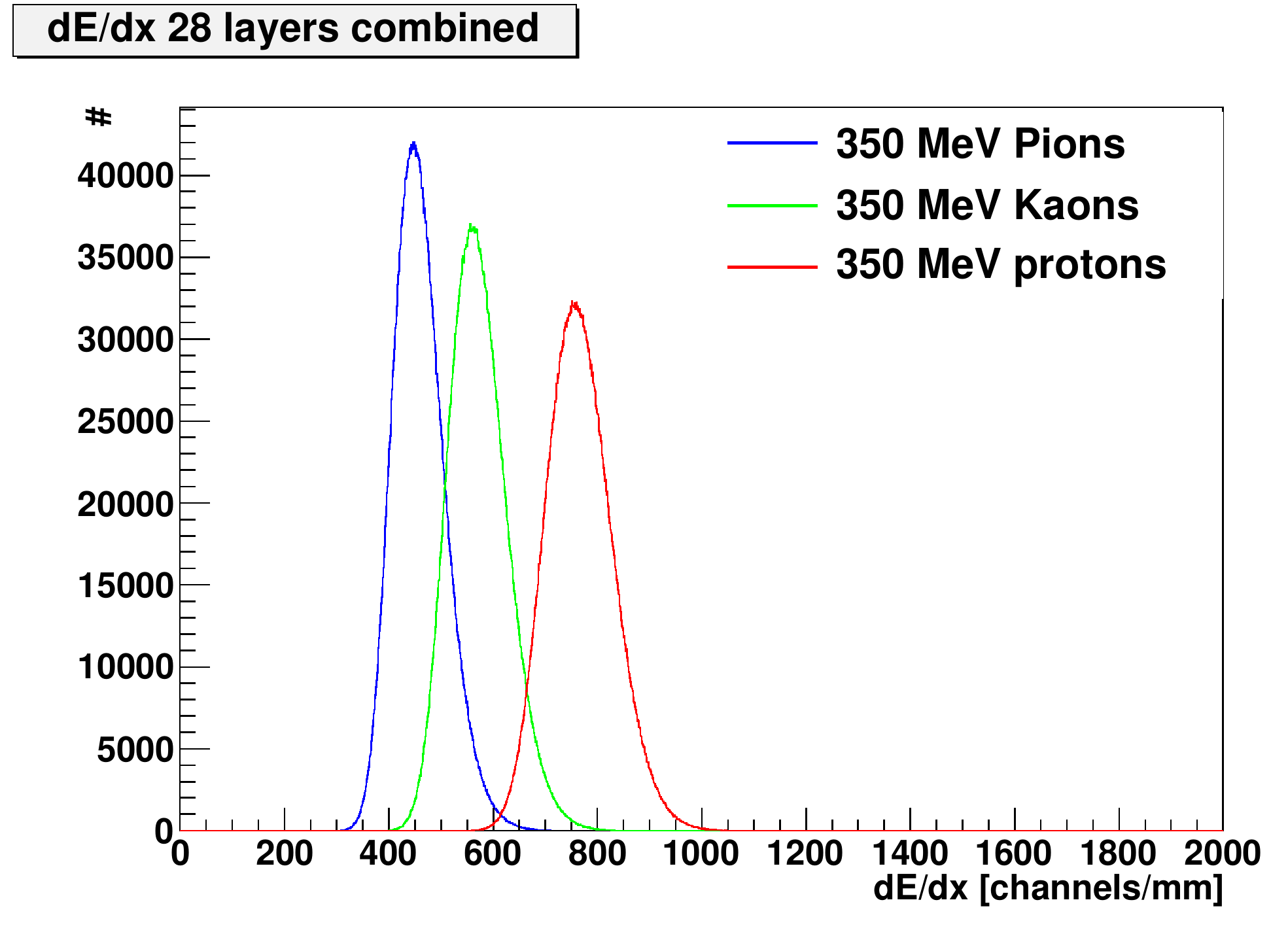}
\end{tabular}
\caption[]{\label{pik350} Pion, kaon, and proton ${\rm d}E/{\rm d}x$ distributions of several layers combined.}
\end{figure}

In Fig.~\ref{350onehit} the simulated energy loss in one straw of pions, kaons, and protons with momenta of 350~MeV is shown. The expected Landau shaped curve is observed. 
Fig.~\ref{pik350} shows the ${\rm d}E/{\rm d}x$ distributions of pions, kaons, and protons with a momentum of 350~MeV for 14 and 28 layers combined. A reasonable separation between pions and protons can be achieved, while it will be difficult to distinguish kaons from pions for example. This PID information can be included in a kinematic fit to identify the most likely particle type for a given track.

\subsubsection{250~MeV Pions and kaons}
\begin{figure}
\begin{tabular}{cc}
  \includegraphics[width=0.95\linewidth]{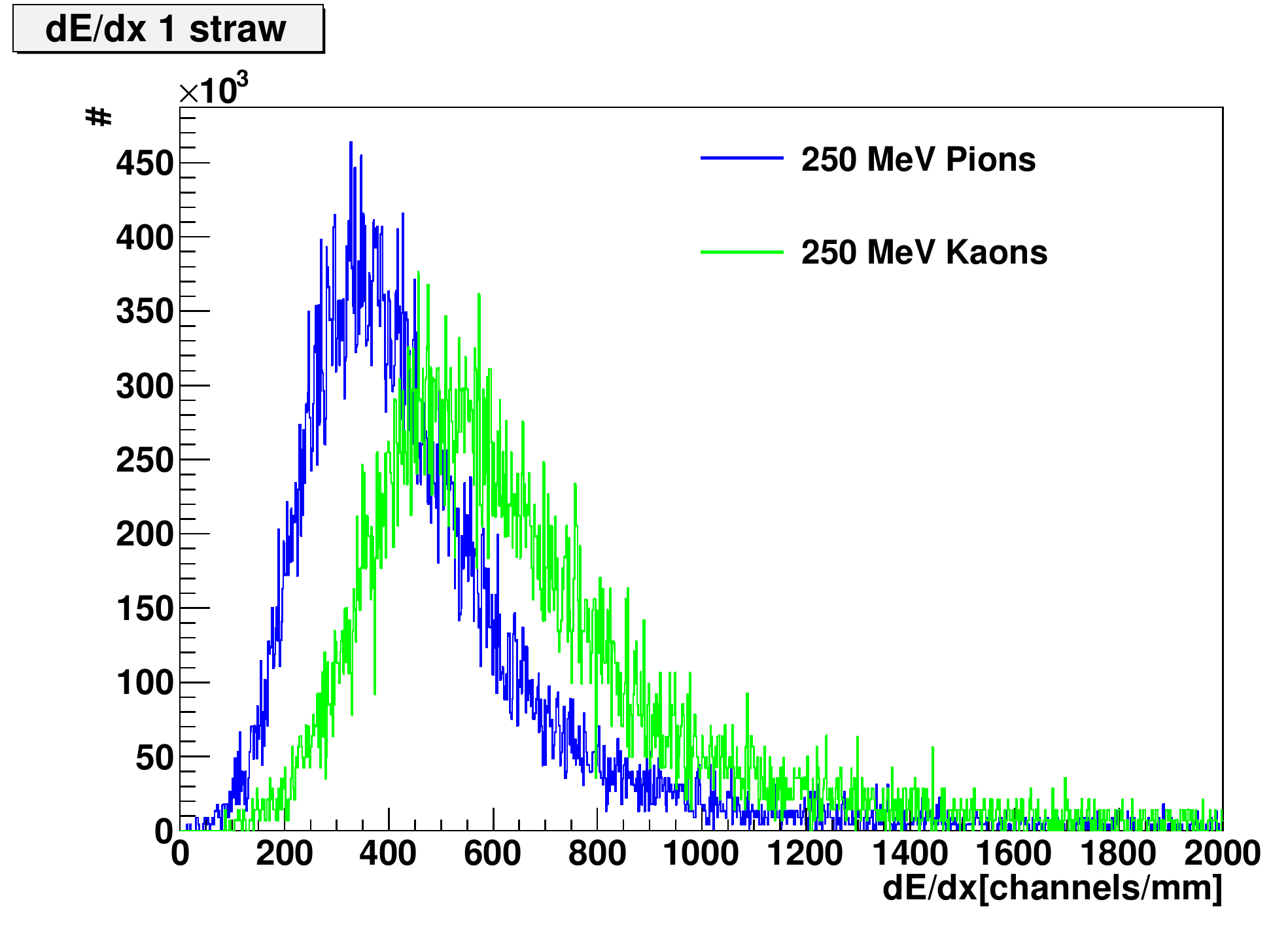}
\end{tabular}
\caption[]{\label{250onehit} Pion and kaon ${\rm d}E/{\rm d}x$ distributions.}
\end{figure}

\begin{figure}
\begin{tabular}{cc}
  \includegraphics[width=0.5\linewidth]{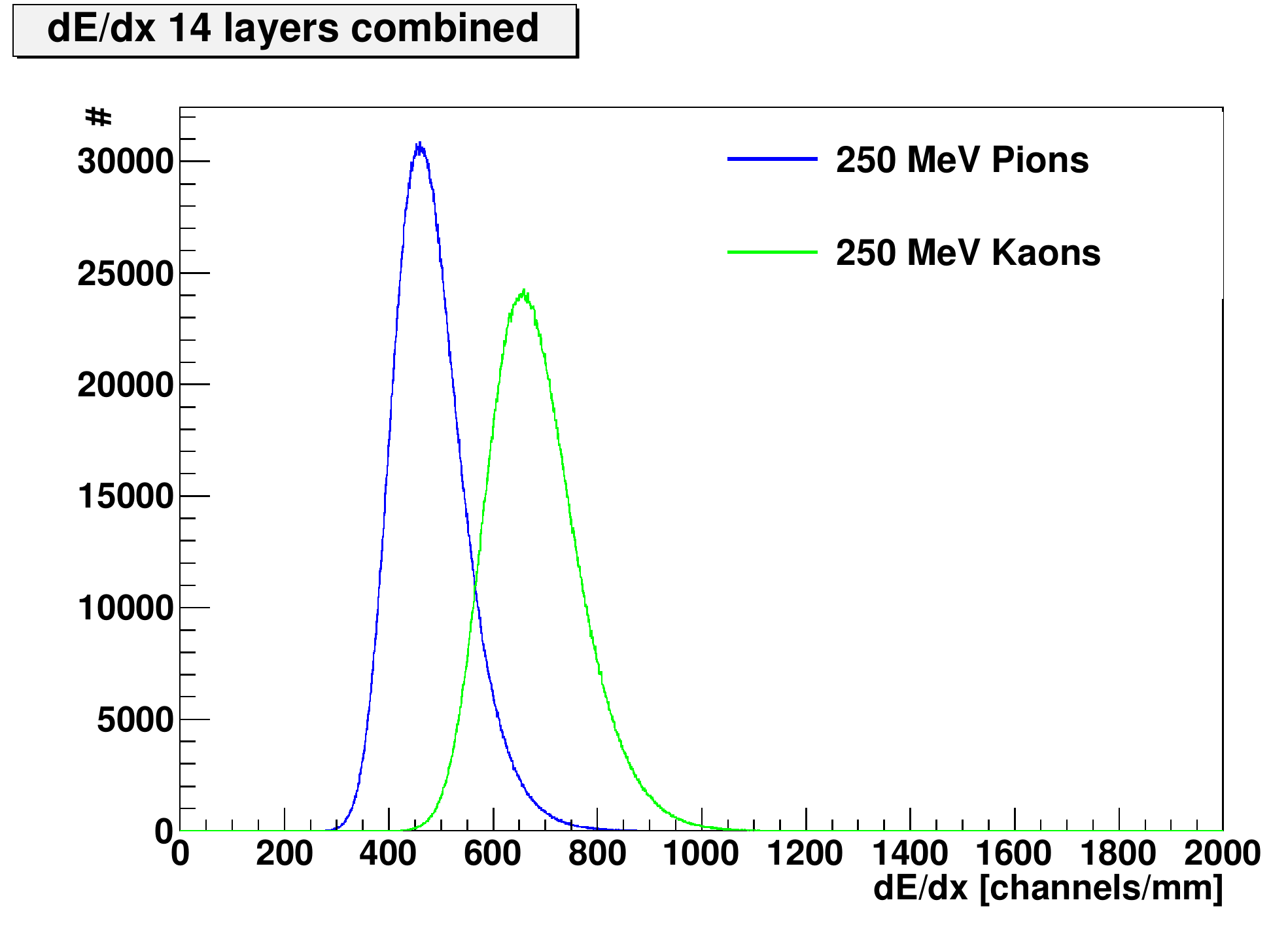}
  \includegraphics[width=0.5\linewidth]{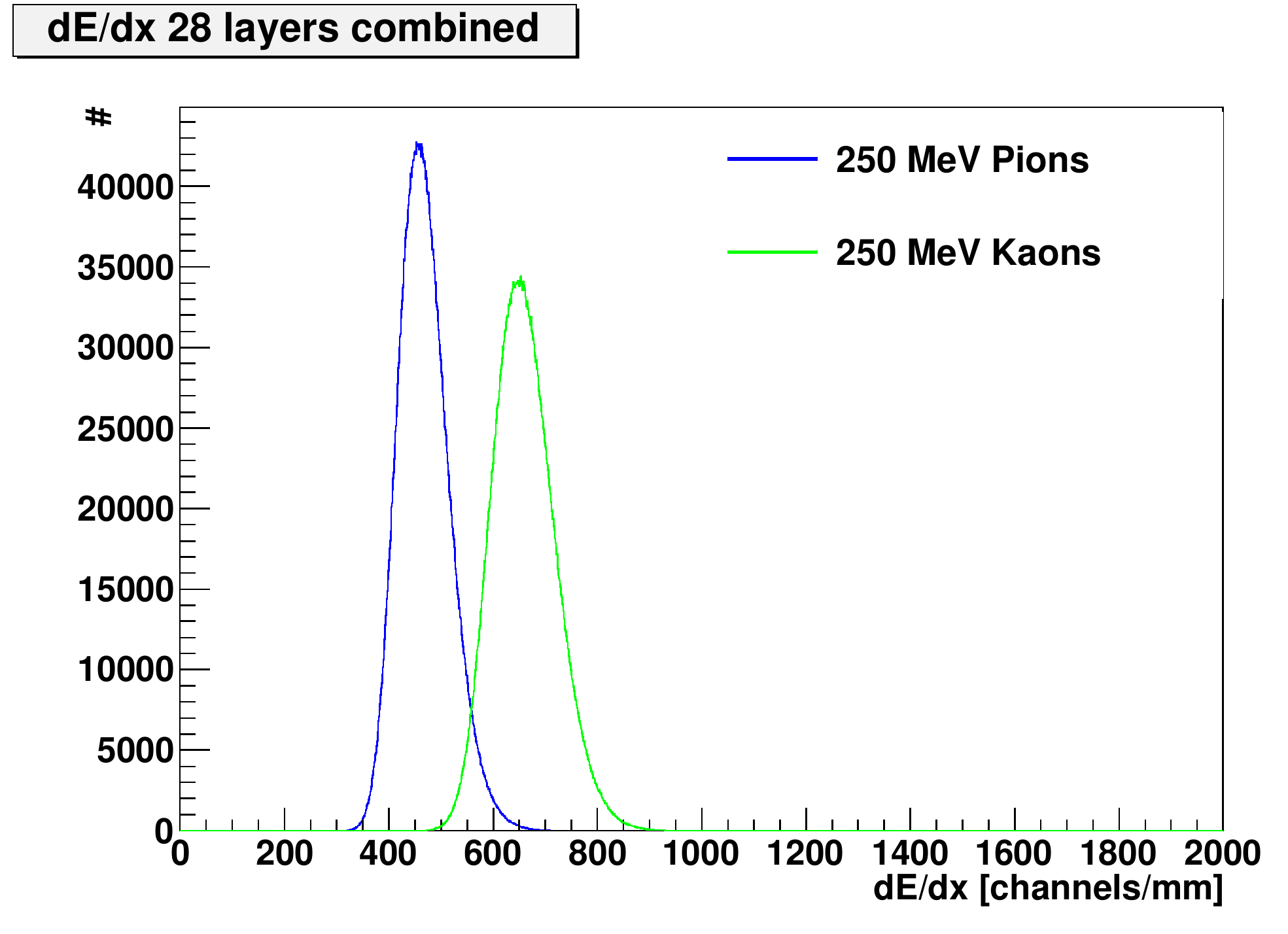}
\end{tabular}
\caption[]{\label{pik250} Pion and kaon ${\rm d}E/{\rm d}x$ distributions of several layers combined.}
\end{figure}

Fig.~\ref{250onehit} shows the energy loss distribution in one straw of pions and kaons with a momentum of 250~MeV. Protons are not shown here because their momentum is not sufficient to pass through more than a few layers of straws.
Fig.~\ref{pik250} shows the energy loss distributions of pions and kaons with a momentum of 250~MeV for several layers combined. A reasonable separation between pions and kaons is observed. 
\section{Charge division}
In this study, the full-scale prototype shown in Fig.~\ref{fullprot} is used. Charge division is established by connecting two neighboring straws with a 60~$\Omega$ resistor at the downstream (far) end of the straw. A two meter long wire has a resistance of 300~$\Omega$. When a signal is induced on a wire the charge generated will be divided between the two connected wires. By taking the ratio of the two signal amplitudes, the position along the wire of the source of the signal can be calculated. This charge division thereby provides complementary information to that obtained from a combined analysis of stereo and axial wires.
\begin{figure}
\begin{tabular}{cc}
  \includegraphics[width=1.0\linewidth]{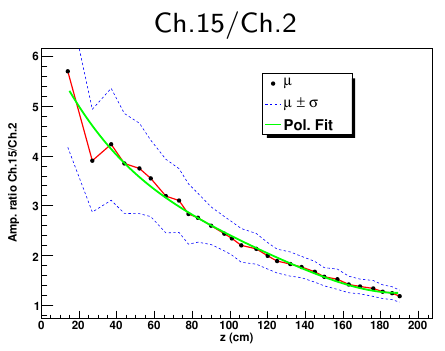}
\end{tabular}
\caption[]{\label{chargediv} Amplitude ratios of $^{55}$Fe induced signals as a function of position along the wire.}
\end{figure}

As a proof of principle, a $^{55}$Fe source was used in a test and placed on straw number 15 that was coupled to straw number 2 by a 60~$\Omega$ junction. With the source positioned at different locations along the wire the amplitudes of the signals were determined and the ratio calculated. The result is shown in Fig.~\ref{chargediv}. In the GlueX experiment, most of the signals are expected to be generated at $z>80$~cm because of the location of the target with respect to the CDC (see introduction). In that region charge division will provide a resolution along the wire of about 10~cm. More details concerning this study can be found in ref.~\cite{chdiv}.
  

\section{Conclusion}
The design of the GlueX straw tube central drift chamber is described in detail. Many design parameters (straw tube material, placement of stereo straws, etc.) have been determined by comparing experimental data from prototype chambers with MC simulation. The choice of gas type and mixture at which to operate the chamber has been determined by properties such as drift-times, which are extracted using flash ADCs, and position resolution. A realistic MC simulation was developed that reproduces the experimental results. The simulation has been used to predict the expected particle identification sensitivity using energy deposition. A test of charge division by connecting neighboring wires has been conducted and shows that position resolution of about 10~cm can be achieved along the direction of the wire.

\section{Acknowledgements}
The authors would like to thank R. Booth (Carnegie Mellon University) for his assistance with using the electron microscope. 

This work was supported in part by the US Department of Energy (under Grant No.
DE-FG02-87ER40315) and  Jefferson Science Associates, LLC operated Thomas Jefferson National Accelerator Facility for the United States Department of Energy under U.S. DOE Contract No. DE-AC05-06OR23177 




\end{document}